\documentclass[12pt]{iopart}
\usepackage{epsfig}
\begin{document}
%==============================================================================================================%
\title[Electron interference phenomena in mesoscopic devices]
{Electron interference in mesoscopic devices in the presence of nonclassical
electromagnetic fields}

\author{D. I. Tsomokos$^{1,2}$, C. C. Chong$^3$, A. Vourdas$^{1}$}
\address{$^1$Department of Computing, University of Bradford,
Bradford, BD7 1DP, England \\
$^2$Department of Physics, Astrophysics and Mathematics, University of Hertfordshire,
Hatfield, AL10 9AB, England \\
$^3$Institute of High Performance Computing, 1 Science Park Road,
117528, Singapore}

\begin{abstract}
The interaction of mesoscopic interference devices with nonclassical
electromagnetic fields is studied. The external quantum fields
induce a phase factor on the electric charges. This phase factor,
which is a generalization of the standard Aharonov-Bohm phase
factor, is in the case of nonclassical electromagnetic fields a
quantum mechanical operator. Its expectation value depends on the
density matrix describing the nonclassical photons and determines
the interference. Several examples are discussed, which show that
the quantum noise of the nonclassical photons destroys slightly the
electron interference fringes. An interesting application arises in
the context of distant electron interference devices, irradiated
with entangled photons. In this case the interfering electrons in
the two devices become entangled. The same ideas are applied in the
context of SQUID rings irradiated with nonclassical electromagnetic
fields. It is shown that the statistics of the Cooper pairs
tunneling through the Josephson junction depend on the statistics of
the photons.
\end{abstract}

 \maketitle
%==============================================================================================================%
%----------------------------------- *** SECTION 1 *** --------------------------------------------------------1
\section{Introduction}
Interference of electrons in the presence of a magnetostatic flux
has been studied for a long time since the work of Aharonov and Bohm
\cite{AB}. The Aharonov-Bohm phase factor is acquired by electric
charges that encircle a magnetic flux, even if the flux vanishes in
the vicinity of the paths of the charges. The effect has inspired
numerous applications in solid state physics \cite{solid_state_AB}.
In particular we mention extensive theoretical and experimental
research on persistent currents in mesoscopic rings
\cite{mesoscopic,1,2,3,4,5}.

Electron interference in the presence of a time-dependent magnetic
flux (i.e., electromagnetic fields) has also been studied
\cite{Vourdas_AB,Vourdas_2001,Chiao,Ford_AB,AC1,AC2}. The intention
here is not to prove the reality of the vector potential, but to
study how electromagnetic fields affect interfering electrons.

The next step in this line of research is to consider nonclassical
electromagnetic fields \cite{FIELDS_1,FIELDS_2} which are carefully
prepared in a particular quantum state, and study their effect on
quantum interference \cite{Vourdas_AB,Vourdas_2001}. In this case it
is shown that the quantum noise in the electromagnetic field
destroys partly the electron interference fringes. Different types
of nonclassical electromagnetic fields are characterized by
different quantum statistics; and we will show explicitly that the
electron interference results depend on the photon statistics.

An important feature of multimode quantum electromagnetic fields is
entanglement. Two electromagnetic field modes can be factorizable
(uncorrelated); or separable (classically correlated); or entangled
(quantum mechanically correlated) \cite{ENT}. Entangled
electromagnetic fields have been produced experimentally in
laboratories for a long time \cite{ENT_FIELDS}. In the context of
this review article, we consider two distant mesoscopic electron
interference devices that are irradiated with a two-mode
nonclassical electromagnetic field \cite{TCV}. Each field mode is
coupled to one of the mesoscopic devices. For entangled
electromagnetic fields, the electric currents in the distant
mesoscopic devices become correlated. Moreover the induced
correlations of the electrons depend on the nature of the
correlation between the external photons.

Similar phenomena can be studied in the context of superconducting
quantum interference devices (SQUID)
\cite{josephson,SQUID_1,SQUID_2,SQUID_3}. Experimental work has so
far concentrated on the interaction of mesoscopic devices with
classical electromagnetic fields. However the interaction of a
Josephson device with a single microwave photon has recently been
studied experimentally in reference \cite{experiment}.

The interaction of mesoscopic SQUID rings with nonclassical
electromagnetic fields has been studied theoretically in
\cite{Jos_Vourdas94, Jos_Vourdas96,SQUID_rings_Hamiltonian,
SQUID_rings}. In this case the Josephson current is a quantum
mechanical operator, whose expectation value with respect to the
density matrix of the external photons, yields the observed current.
The interaction of entangled electromagnetic fields with two
spatially separated SQUID rings has been studied in
\cite{TCV_SQUID,two_distant_SQUID}. It has been shown that the
photon correlations can be transferred to the Cooper pair currents
measured in the two distant SQUID rings.

In this interdisciplinary work we bridge the gap between electron
coherence in mesoscopic physics and nonclassical phenomena in
quantum optics. Work on entanglement of several mesoscopic devices
has been reported in \cite{ENT_mesoscopic_devices}.

The paper is organized as follows. In section 2 we describe certain
one-mode and two-mode nonclassical fields, which are relevant to the
rest of our work, and derive the corresponding Weyl function
\cite{Weyl_function}. In section 3 we discuss the magnetic flux and
the electromotive force operators, which are the dual quantum
variables in our context.

We subsequently turn our attention to electron interference
phenomena. In section 4 we describe the standard Aharonov-Bohm phase
factor in electron interference that is induced by a magnetostatic
flux. In section 5 we describe the electron phase factor
operator that is induced by nonclassical electromagnetic fields
\cite{Vourdas_AB}. It is explained that the expectation value of the
phase factor and, consequently, of the electron intensity
distribution depend on the quantum state of the external photons.

We stress that accurate knowledge of the quantum state of the
electromagnetic field enables us to calculate not only the average
intensity of the interfering electrons, but also their full
statistics (higher order correlations). In section 6 we quantify the
quantum statistics of the interfering electrons using the
autocorrelation function and its Fourier transform, the spectral
density. It is shown that the quantum statistics of the interfering
electrons depend on the quantum statistics of the photons
\cite{Vourdas_2001}. In section 7 we describe how two spatially
separated electron interference experiments, which interact with
entangled fields, become correlated \cite{TCV}.

In section 8 we study the interaction of nonclassical
electromagnetic fields with mesoscopic SQUID rings. In the case of
two distant SQUID rings, which are coupled to two entangled
electromagnetic fields, we show that the quantum currents tunneling
through the distant Josephson junctions become entangled
\cite{TCV_SQUID}. We conclude with a summary of the results in
section 9.

%==============================================================================================================%
%----------------------------------- *** SECTION 2 *** --------------------------------------------------------2
\section{Nonclassical electromagnetic fields}
In this section we introduce the nonclassical states of the
electromagnetic field that are relevant to the rest of our work. We
define the Weyl function and provide its value in the case of
number, coherent, squeezed, and thermal states. It is noted that we
use theoretical units, in which $k_{\rm B} = \hbar = c =1$.

\subsection{One-mode quantum states of the electromagnetic field}
Nonclassical electromagnetic fields are carefully prepared in a
particular quantum state and are described by a density matrix
$\rho$. In this case we know the average values $\langle E \rangle,
\langle B \rangle$ of the electric and magnetic fields, the standard
deviations $\Delta E,\Delta B$ and also their higher moments.
Another quantity which describes the fields is the photon counting
distribution function
\begin{equation}\label{P_N}
P(N)\equiv \langle N |\rho |N \rangle.
\end{equation}
Various examples of nonclassical electromagnetic fields are given
below.

\subsubsection{Number states}
The number states $|N\rangle$ are defined as:
\begin{eqnarray} \label{number_state}
|N\rangle=\frac{(\hat{a}^{\dag})^{N}}{\sqrt{N!}}|0\rangle.
\end{eqnarray}

\subsubsection{Coherent states}
The coherent states $|A\rangle$ are defined as:
\begin{eqnarray}\label{coherent_state}
|A \rangle = D(A) |0 \rangle
\end{eqnarray}
where $D(A)$ is the displacement operator
\begin{eqnarray}\label{displacement_operator}
D(z)=\exp(z \hat{a}^{\dag}-z^{*} \hat{a}).
\end{eqnarray}
The photon counting distribution is in this case Poissonian.

\subsubsection{Squeezed states}
The squeezing operator is defined as
\begin{eqnarray} \label{squeezing_operator}
S(r\varphi) = \exp\left[-\frac{r}{4}\exp(-i\varphi)\hat{a}^{\dagger 2}
+\frac{r}{4}\exp(i\varphi)\hat{a}^{2}\right]
\end{eqnarray}
where the $r,\varphi$ are real numbers and $r$ is known as the
squeezing parameter. Squeezed states $|A;r\varphi \rangle $ are
defined by acting on the coherent state $|A\rangle$, with the
squeezing operator
\begin{eqnarray} \label{squeezed_states}
|A;r\varphi \rangle =S(r\varphi) |A\rangle = S(r\varphi)D(A)|0\rangle .
\end{eqnarray}
In this case $P(N)$ can be sub-Poissonian. The average number of photons is
\begin{eqnarray} \label{squeezed_average_N}
\langle N \rangle _{\rm sq} = \left[\sinh
\left(\frac{r}{2}\right)\right]^2+\left[\cosh \left(\frac{r}{2}\right) -
\sinh\left(\frac{r}{2}\right) \right]^2 |A|^2.
\end{eqnarray}

In figure 1 we have plotted the electric field as a function of time
in the case of coherent and squeezed light. Both the average value
$\langle \hat E \rangle$ and the quantum noise $\Delta \hat E$ are
shown. The parameters are chosen so that the average value of the
electric field is the same in both examples. It is seen that the two
fields differ in the quantum noise $\Delta \hat E$. The
anti-bunching of photons in squeezed states, in comparison to the
Poissonian statistics in the case of coherent states, is also shown
in the figure. These two types of nonclassical electromagnetic
fields will be used later, in the context of electron interference
(i.e., we will study the situation where these nonclassical
electromagnetic fields are coupled with electron interference
devices). It will be shown there that they produce different results
for the electron interference.

%======================== FIGURE 1
\begin{figure} [ht]
\centering \scalebox{0.55}
{\includegraphics{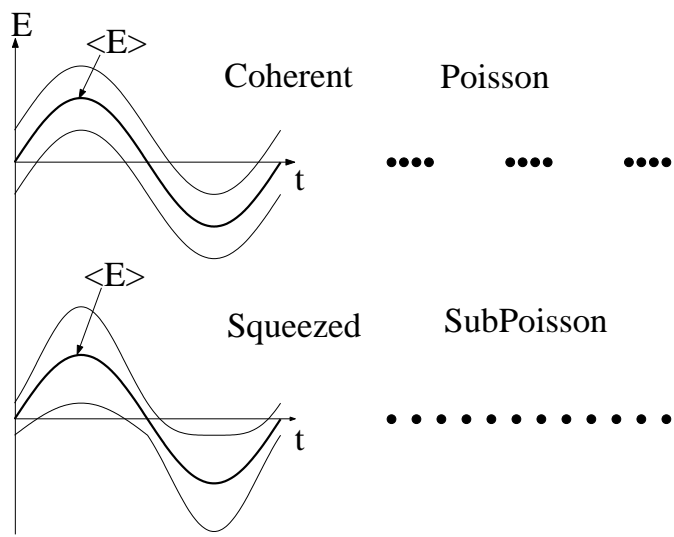}}\caption{The electric field of
coherent and squeezed light as a function of time. Both the average
value $\langle \hat E \rangle$ and the quantum noise $\Delta \hat E$
are shown. The anti-bunching of photons in squeezed states, in
comparison to the Poissonian statistics in coherent states, is also
shown.}
\end{figure}

\subsubsection{Thermal states}
The thermal states are defined through the density matrix
\begin{eqnarray} \label{thermal_states}
\rho_{\rm th} &=& \left[1-\exp(-\beta \omega)\right]\exp(-\beta
\omega\hat{a}^{\dag}\hat{a}) \nonumber \\
&=& \left[1-\exp(-\beta \omega)\right] \sum_{n=0}^{\infty} \exp(-\beta\omega
n)|n\rangle \langle n |
\end{eqnarray}
where $\beta$ is the inverse temperature. In this case the average
number of photons is
\begin{eqnarray} \label{thermal_average_N}
\langle N \rangle _{\rm th} = \frac{1}{\exp(\beta \omega) -1}.
\end{eqnarray}

\subsection{Weyl functions}
The Wigner and Weyl (or characteristic) functions play an important
role in quantum mechanics \cite{Weyl_function}. The Weyl function
that corresponds to a quantum state described by a density operator
$\rho$ is defined in terms of the displacement operator of equation
(\ref{displacement_operator}) as
\begin{eqnarray} \label{Weyl_function}
\tilde {W} (z) \equiv \Tr [\rho D(z)].
\end{eqnarray}
The tilde in the notation reflects the fact that the Weyl function
$\tilde {W}$ is the two-dimensional Fourier transform of the Wigner
function $W$. The $\tilde {W} (z)$ is a complex function, in
general, whose absolute value obeys
\begin{eqnarray}\label{abs_Weyl}
0 \le |\tilde {W} (z)| \le 1.
\end{eqnarray}

For later use we give the Weyl function for various states. We start
with the following relation \cite{Roy}
\begin{eqnarray} \label{matrix_elements_D}
\langle M |D(z)|N \rangle = \left(\frac{N!}{M!}\right)^{1/2} z^{M-N} \exp
\left(-\frac{|z|^2}{2}\right) L_{N}^{M-N}(|z|^2)
\end{eqnarray}
where the $L_{k}^{\alpha}$ are Laguerre polynomials \cite{tables}.
Therefore the Weyl function for a number state $|N\rangle$ is
\begin{eqnarray} \label{Weyl_number}
\tilde{W}_{\rm num}(z)=\exp\left(- \frac{|z|^2}{2}\right) L_{N}(|z|^2).
\end{eqnarray}

The Weyl function for a coherent state $|A\rangle$ is
\begin{eqnarray} \label{Weyl_coherent}
\tilde{W}_{\rm coh} (z) = \exp \left[- \frac{|z|^2}{2}+i2|A z|\sin(\arg{z}
-\arg{A})\right].
\end{eqnarray}

The Weyl function for a squeezed state $|A;r\varphi \rangle$ is
\begin{eqnarray} \label{Weyl_squeezed}
\fl \tilde{W}_{\rm sq}(z) =  \exp(-Y+iX), \\
\fl X = 2|A z| \left[\cosh\left(\frac{r}{2}\right)\sin(\arg{z}
-\arg{A})-\sinh\left(\frac{r}{2}\right) \sin(\arg{z} + \arg{A}
+\varphi)\right], \nonumber \\
\fl Y = \frac{|z|^2}{2}\left[\cosh(r)+\sinh(r)\cos(2\arg{z}+\varphi)\right].
\nonumber
\end{eqnarray}

Finally for thermal states we have
\begin{equation} \label{Weyl_thermal}
\tilde{W}_{\rm th}(z=\zeta e^{i\omega t})=\exp\left[-\frac{\zeta^2}{2}
\coth\left(\frac{\beta\omega}{2}\right)\right],
\end{equation}
where $\zeta$ is a real number. These relations have been given in reference
\cite{Jos_Vourdas94}.

\subsection{Two-mode quantum states: separability versus entanglement}
Nonclassical electromagnetic fields with several modes allow for
correlations between the distinct field modes. The nature of the
correlation can be classical or quantum \cite{ENT}.

Let $\rho$ be the density matrix that describes a two-mode
nonclassical electromagnetic field. Then the density matrices of the
two fields are
\begin{eqnarray}\label{rho_A}
\rho_{\rm A}\equiv {\rm Tr}_{\rm B} (\rho),  \;\;\;\;\;\; \rho_{\rm B}\equiv
{\rm Tr}_{\rm A} (\rho).
\end{eqnarray}
The density matrix $\rho$ for the two-mode electromagnetic field
state is factorizable if $\rho_{\rm fact} = \rho_{\rm A}\otimes
\rho_{\rm B}$. The density matrix $\rho$ is separable if
\begin{eqnarray} \label{rho_sep}
\rho_{\rm sep} = \sum_k P_k \rho_{{\rm A},k} \otimes \rho_{{\rm B},k}
\end{eqnarray}
where $P_k$ are probabilities. In all other cases the density matrix
$\rho_{\rm ent}$ is entangled.

\subsubsection{Two-mode number states}
For later use we consider the (mixed) separable density operator
\begin{eqnarray} \label{rho_sep_num}
\rho_{\rm sep} = \frac{1}{2}(|N_1 N_2 \rangle \langle N_1 N_2| + |N_2 N_1
\rangle \langle N_2 N_1 |).
\end{eqnarray}
We also consider the (pure) entangled state $|s \rangle =
{2}^{-1/2}(|N_1 N_2 \rangle + |N_2 N_1 \rangle)$, for example. The
corresponding density operator is
\begin{eqnarray}\label{rho_ent_num}
\rho_{\rm ent} = \rho_{\rm sep} + \frac{1}{2}(|N_1 N_2 \rangle \langle N_2
N_1| + |N_2 N_1 \rangle \langle N_1 N_2|).
\end{eqnarray}
Clearly in this example the $\rho_{\rm sep}$ and the $\rho_{\rm
ent}$ differ only in the above nondiagonal elements. In both the
separable and the entangled case the reduced density operators of
equation (\ref{rho_A}) are given by
\begin{eqnarray}\label{reduced_rho_num}
\rho_{\rm sep, A} = \rho_{\rm ent, A}=\rho_{\rm sep, B}=\rho_{\rm ent, B} =
\frac{1}{2} (|N_1\rangle \langle N_1| + |N_2\rangle\langle N_2|).
\end{eqnarray}

\subsubsection{Two-mode coherent states}
We consider the two-mode coherent states in the classically
correlated state
\begin{eqnarray}\label{rho_sep_coherent}
\rho_{{\rm sep}}=\frac{1}{2}(|A_1 A_2\rangle \langle A_1 A_2| +|A_2 A_1\rangle
\langle A_2 A_1|).
\end{eqnarray}
In this case the reduced density operators are
\begin{eqnarray}\label{reduced_rho_coh_sep}
\rho_{\rm sep,A}=\rho_{\rm sep,B}=\frac{1}{2}(|A_1\rangle \langle A_1|
+|A_2\rangle\langle A_2|).
\end{eqnarray}

We also consider the entangled state $|s \rangle={\cal N}(|A_1
A_2\rangle +|A_2 A_1\rangle)$ with density operator
\begin{eqnarray}\label{rho_ent_coherent}
\rho_{\rm ent}= 2{\cal N} ^2 \rho_{\rm sep}+ {\cal N}^2 (|A_1 A_2\rangle
\langle A_2 A_1| +|A_2 A_1\rangle \langle A_1 A_2|)
\end{eqnarray}
where the normalization constant is given by
\begin{eqnarray}\label{normalization_constant}
{\cal N}=\left[2+2\exp\left(-|A_1-A_2|^2\right)\right]^{-1/2}.
\end{eqnarray}
In this case the reduced density operators are
\begin{eqnarray}\label{reduced_rho__coh_ent}
\fl \rho_{\rm ent,A} = \rho_{\rm ent,B}= {\cal N}^2(|A_1\rangle \langle A_1|
+|A_2\rangle \langle A_2| + \chi |A_1\rangle\langle A_2| + \chi ^{*}
|A_2\rangle\langle A_1|)
\end{eqnarray}
where
\begin{eqnarray} \label{chi}
\chi = \langle A_1|A_2\rangle = \exp\left(-\frac{|A_1|^2}{2}
-\frac{|A_2|^2}{2} + A_1^{*} A_2 \right).
\end{eqnarray}

%==============================================================================================================%
%----------------------------------- *** SECTION 3 *** --------------------------------------------------------3
\section{Magnetic flux operator}
We consider a monochromatic electromagnetic field of frequency
$\omega$, at sufficiently low temperatures $T \ll \omega$, so that
the quantum noise is greater than the thermal noise. In this case
the vector potential $\hat{A}_i$ and the electric field $\hat{E}_i$
are dual quantum variables. For a loop $C$, which is small in
comparison to the wavelength of the electromagnetic field, the
$\hat{A}_i, \hat{E}_i$ are integrated around $C$ and yield the
magnetic flux $\hat{\phi} =\oint _C \hat{A}_i dx_i$ and the
electromotive force $\hat{V}_{\rm EMF}=\oint _C \hat{E}_i dx_i$,
correspondingly, as dual quantum variables.

In terms of these variables the photon creation and annihilation
operators are
\begin{eqnarray}\label{creation__destruction_operators}
\hat a^{\dag}=\frac{1}{\sqrt{2}\xi}(\hat\phi-i\omega^{-1}\hat V_{{\rm EMF}}),
\;\;\;\;\; \hat{a}=\frac{1}{\sqrt{2}\xi}\left(\hat{\phi}+i\omega^{-1}
\hat{V}_{\rm EMF}\right),
\end{eqnarray}
where $\xi$ is a constant proportional to the area enclosed by $C$.
Consequently the magnetic flux operator is $\hat{\phi} (0)=
2^{-1/2}\xi(\hat{a}^\dagger + \hat{a})$ and its evolution in time is
given by
\begin{eqnarray} \label{flux_op_Heisenberg_picture}
\hat {\phi}(t)= \exp (it{\cal H})\hat{\phi}(0)\exp (-it{\cal H}).
\end{eqnarray}
The Hamiltonian ${\cal H}$ of the system is
\begin{eqnarray} \label{free_Hamiltonian}
{\cal H} =\omega(a^{\dagger}a+1/2) + {\cal H}_{\rm int}.
\end{eqnarray}
In the external field approximation we ignore the interaction
Hamiltonian ${\cal H}_{\rm int}$ and we obtain
\begin{eqnarray} \label{quantum_flux}
\hat \phi(t)=\frac{\xi}{\sqrt{2}} \left[\exp(i\omega t)\hat{a}^\dagger +
\exp(-i\omega t)\hat{a}\right].
\end{eqnarray}
This is a good approximation when the flux due to back reaction is
small in comparison to the external flux.

The expectation value of the flux and the quantum uncertainty
$\Delta \hat \phi $ are given by
\begin{eqnarray} \label{exp_value_phi}
\langle \hat \phi(t) \rangle = \Tr [\rho \hat \phi(t)], \;\;\;\;\; \Delta \hat
\phi = [\langle \hat \phi^{2}(t) \rangle - \langle \hat \phi(t) \rangle
^{2}]^{1/2}.
\end{eqnarray}
For example, in the case of number states $\rho = |N \rangle \langle
N|$ we get
\begin{eqnarray} \label{exp_value_phi_num}
\langle \hat \phi(t) \rangle _{\rm num} = 0, \;\;\;\;\; (\Delta \hat \phi)
_{\rm num} = \left(N+\frac{1}{2}\right)^ {1/2}.
\end{eqnarray}
For coherent states $\rho = | A\rangle \langle A|$ we have
\begin{eqnarray} \label{exp_value_phi_coh}
\langle \hat \phi(t) \rangle _{\rm coh} = 2^{1/2}|A|\cos(\omega t - \arg{A}),
\;\;\;\;\; (\Delta \hat \phi) _{\rm coh} = 2^{-1/2}.
\end{eqnarray}
In the case of squeezed states $\rho = |A; r\varphi \rangle \langle
A; r\varphi|$ we obtain
\begin{eqnarray} \label{exp_value_phi_sq}
\langle \hat \phi(t) \rangle _{\rm sq} = - 2^{1/2}|A||Z| \cos(\omega t +
\arg{Z}), \\
(\Delta \hat \phi) _{\rm sq} = 2^{-1/2} [\cosh(r)-\sinh(r)
\cos(2\omega t + \varphi)]^{1/2}, \nonumber \\
Z = \sinh(r/2)\exp[i(\arg{A}+\varphi)] - \cosh(r/2) \exp(-i\arg{A}). \nonumber
\end{eqnarray}
Finally for the thermal states of equation (\ref{thermal_states}) we
get
\begin{eqnarray} \label{exp_value_phi_th}
\langle \hat \phi(t) \rangle _{\rm th} = 0, \;\;\;\;\; (\Delta \hat \phi)
_{\rm th} = \frac{1}{\sqrt{2}} \left[\coth \left( \frac{1}{2} \beta \omega
\right) \right]^{1/2}.
\end{eqnarray}

%==============================================================================================================%
%----------------------------------- *** SECTION 4 *** --------------------------------------------------------4
\section{Aharonov-Bohm phase factor induced by a magnetostatic flux}
We consider a two-path electron interference experiment, as shown in
figure 2.

%======================== FIGURE 2
\begin{figure} [ht]
\centering \scalebox{0.55}
{\includegraphics{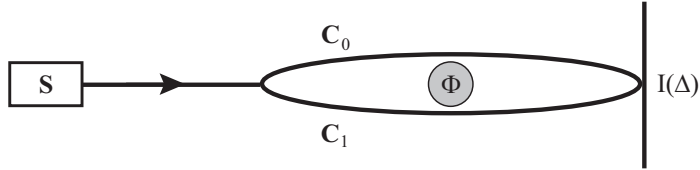}}\caption{Aharonov-Bohm experiment. The
electrons follow the lowest winding paths $C_0,C_1$ in a field-free region.
The loop $C_0-C_1$ is threaded by a magnetostatic flux $\Phi$.}
\end{figure}

The wavefunctions corresponding to paths $C_0$ and $C_1$ are
$\psi_0$ and $\psi_1$, respectively. In the presence of magnetic
flux $\Phi$ threading the loop $C_0-C_1$, we get the electron
intensity
\begin{eqnarray}\label{mitsos}
I(x)=|\psi_0|^2+|\psi_1|^2+2|\psi_0 \psi_1|\cos(x - e\Phi),
\end{eqnarray}
where $x$ is the phase difference between the two paths:
\begin{eqnarray}\label{x_definition}
x (\Delta) \equiv \arg(\psi_0)-\arg(\psi_1).
\end{eqnarray}
If we assume equal splitting (i.e., $|\psi_0|^2=1/2=|\psi_1|^2$) then
\begin{eqnarray} \label{intensity}
I(x)=1+\cos(x - e\Phi).
\end{eqnarray}
The visibility of the intensity $I$, defined as
\begin{eqnarray}\label{visibility}
\nu=\frac{I_{\rm max}-I_{\rm min}}{I_{\rm max}+I_{\rm min}},
\end{eqnarray}
is equal to one in this case.

%==============================================================================================================%
%----------------------------------- *** SECTION 5 *** --------------------------------------------------------5
\section{Phase factor operator induced by nonclassical electromagnetic fields}
In this section we consider a mesoscopic electron interference
device (${\rm \sim 0.1 \mu m}$) in a microwave waveguide at low
temperatures (${\rm 10-100 mK}$). The electric field is parallel to
the plane of the electron paths and the magnetic field is
perpendicular to it (figure 3). The electron intensity is given by
equation (\ref{mitsos}) where the flux $\Phi$ is now time-dependent.

We next consider the case where the microwaves are nonclassical. In
this case $\Phi$ is a quantum mechanical operator and its
expectation value with respect to the density matrix $\rho$ of the
microwaves gives the observed electron intensity:
\begin{eqnarray} \label{intensity1}
\fl I(x,t)=1+{\rm Tr} [\rho \cos(x - e{\hat \phi})]=1+{\Re}[{\rm
e}^{ix}{\tilde W}(\lambda)] =1 + |\tilde{W}(\lambda)| \cos\{ x -
\arg[\tilde{W}(\lambda)] \}.
\end{eqnarray}
Here ${\tilde W}$ is the Weyl function of the density matrix $\rho$
defined in equation (\ref{Weyl_function}), and we define
\begin{eqnarray} \label{q_definition}
\lambda = iq\exp(i\omega t), \;\;\;\;\;\;q=\frac{\xi e}{\sqrt{2}}.
\end{eqnarray}
If we compare and contrast equation (\ref{intensity}) for classical
microwaves, with equation (\ref{intensity1}) for nonclassical
microwaves we see that the visibility is reduced in the second case
from $1$ to $|\tilde{W}(\lambda)|$. This is due to the quantum noise
in the nonclassical microwaves as can be seen from the expansion
\begin{eqnarray}
\fl |\tilde{W}(\lambda)|^2=1-\frac{q^2}{2}[(\Delta X)^2+ (\Delta P)^2]
-\frac{q^2}{2}[(\Delta X)^2 - (\Delta P)^2] \cos(2\omega t) - \ldots
\end{eqnarray}
where $X=\xi^{-1} \hat \phi$, $P=(\omega \xi)^{-1} \hat{V}_{\rm EMF}$, and
$\Delta X, \Delta P$ are the corresponding uncertainties. The
$|\tilde{W}(\lambda)|$ is less than $1$ due to the non-zero values of the
quantum noise $\Delta X, \Delta P$.

Results are given below for the electron intensity $I(x,t)$ that
corresponds to irradiation with several quantum states. We choose
the point $x=0$, for simplicity.

%======================== FIGURE 3
\begin{figure} [ht]
\centering \scalebox{0.35} {\includegraphics{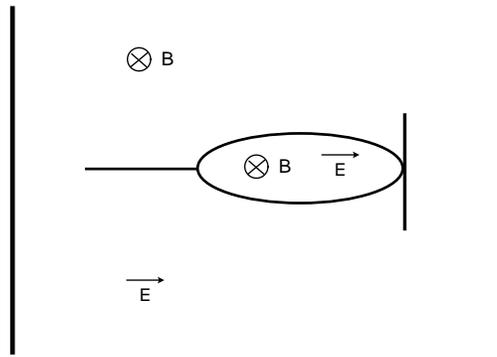}}\caption{Modified
Aharonov-Bohm experiment in the presence of an electromagnetic field. The
field travels in the waveguide with the magnetic field perpendicular to the
plane of the electron paths $C_0,C_1$ and the electric field parallel to it.}
\end{figure}

%======================== FIGURE 4
\begin{figure} [ht]
\centering \scalebox{0.5}
{\includegraphics{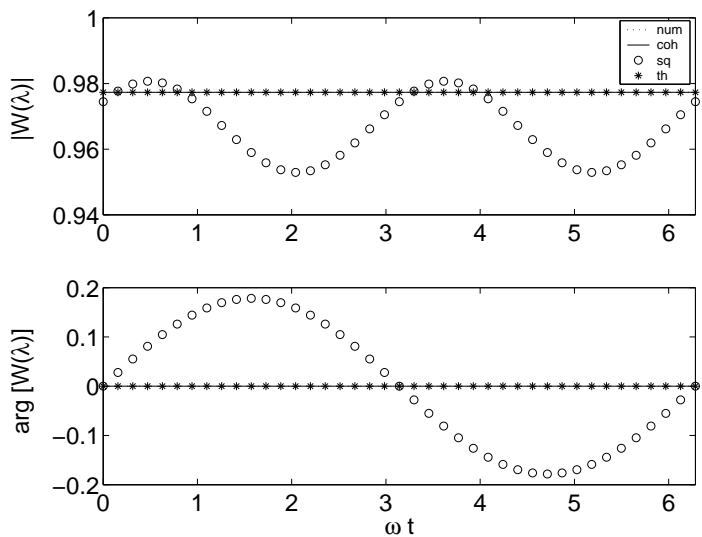}}\caption{Vacuum-induced phase factor for the
charges as a function of time, $\omega t$, for $\omega=10^{-4}$, corresponding
to the case of number states (broken line), coherent states (solid line),
squeezed states (line of circles), and thermal states (line of stars). The
average number of photons $\langle N \rangle$ is zero; the squeezing parameter
$r$ is 0.5. In subplot (a) the $|\tilde {W} (\lambda)|$ is shown and in (b)
the $\arg[\tilde {W} (\lambda)]$ is shown.}
\end{figure}

\subsubsection{Number states}
For the number states of equation (\ref{number_state}), using
equation (\ref{Weyl_number}), we get
\begin{eqnarray} \label{intensity_num}
I_{\rm num}(t) = 1 + \exp \left(-\frac{q^2}{2}\right) L_{N}\left(q^{2}\right).
\end{eqnarray}

\subsubsection{Coherent states}
For the coherent states of equation (\ref{coherent_state}), using
equation (\ref{Weyl_coherent}), we get
\begin{eqnarray} \label{intensity_coh}
I_{\rm coh}(t) = 1 + \exp\left(-\frac{q^2}{2}\right)
\cos\left[2q|A|\cos(\omega t-\arg{A})\right].
\end{eqnarray}
In this case the result is very similar to the classical result of
equation (\ref{intensity}) but the visibility is slightly reduced
from $1$ to $\exp\left(-\frac{q^2}{2}\right)$. The quantum noise of
the coherent states slightly destroys the interference and reduces
its visibility. Even in the absence of microwaves (vacuum state) we
get a reduction in the visibility due to the vacuum noise.

In figure 4 we have plotted the expectation value of the phase
factor operator, which is induced by the electromagnetic vacuum,
the coherent states of equation
(\ref{coherent_state}), the squeezed states of equation
(\ref{squeezed_states}) for $r=0.5$, and the thermal states of
equation (\ref{thermal_states}).

\subsubsection{Squeezed states}
For the squeezed states of equation (\ref{squeezed_states}), using
equation (\ref{Weyl_squeezed}), we get
\begin{eqnarray} \label{intensity_sq}
\fl I_{\rm sq}(t) &=& 1 + \exp(-Y_1)\cos(X_1), \\
\fl Y_1 &=& \frac{q^2}{2}[\cosh(r)-\sinh(r)\cos(2\omega t +\varphi)], \nonumber \\
\fl X_1 &=& 2q|A|\left[\cosh\left(\frac{r}{2}\right)\cos(\omega t -\arg{A}) -
\sinh\left(\frac{r}{2}\right)\cos(\omega t +\arg{A} + \varphi) \right].
\nonumber
\end{eqnarray}
We note that in the case of squeezed vacuum ($A=0$) the intensity
$I_{\rm sq}(t)$ contains all the frequencies $2K\omega$ where $K$ is
an integer (after a Fourier expansion). In contrast in the case of
coherent states we get all the frequencies $K\omega$. The factor of
$2$ in the case of squeezed vacuum is related with the fact that the
squeezed vacuum is a superposition of even number states only.
Therefore the electrons can only absorb an even number of photons
(there are no odd number states in this quantum state). In this case
the result is qualitatively different from the classical result.

\subsubsection{Thermal states}
For the thermal states of equation (\ref{thermal_states}), using
equation (\ref{Weyl_thermal}), we have
\begin{eqnarray} \label{intensity_th}
I_{\rm th}(t) = 1 + \exp\left[-\frac{q^2}{2}
\coth\left(\frac{\beta\omega}{2}\right)\right].
\end{eqnarray}

%==============================================================================================================%
%----------------------------------- *** SECTION 6 *** --------------------------------------------------------6
\section{Quantum statistics of the interfering electrons}
There are various quantities that can be used to describe the
quantum statistics of the interfering electrons. In the previous
section we studied the electron intensity and here we consider
higher order correlations \cite{Vourdas_2001}. We compare and
contrast the results for the two cases of classical and nonclassical
microwaves.

\subsection{Autocorrelation function of the electron
intensity in the case of classical microwaves} In general for a
function $I(t)$ the autocorrelation function is defined as
\begin{eqnarray} \label{autocorr}
\Gamma(\tau) = \lim_{T\rightarrow \infty} \frac{1}{2T} \int_{-T}^{T}
I^{*}(t)I(t+\tau) dt.
\end{eqnarray}
The following properties are well known (e.g., see reference
\cite{Mandel_Wolf}):
\begin{eqnarray} \label{Gamma_properties}
\Gamma(-\tau)=\Gamma^{*}(\tau),\;\;\;\;\; \Gamma(0)\geq 0, \;\;\;\; \;
|\Gamma(\tau)|\leq \Gamma(0).
\end{eqnarray}
The normalized autocorrelation function is defined as
\begin{eqnarray} \label{normalG}
\gamma(\tau) \equiv \frac{\Gamma(\tau)}{\Gamma(0)}, \;\;\;\;\;
0\leq|\gamma(\tau)|\leq1.
\end{eqnarray}
An expansion of $\Gamma(\tau)$ into a Fourier series yields the
spectral density coefficients
\begin{eqnarray} \label{psd}
S_K &=& \frac{\Omega}{2\pi}\int_{0}^{2\pi/\Omega} \Gamma(\tau)
\exp(-iK\Omega\tau) d\tau \\
\Gamma(\tau) &=&\sum_{K=-\infty}^{\infty}S_K\exp(iK\Omega\tau). \nonumber
\end{eqnarray}
The property $\Gamma(-\tau)=\Gamma^{*}(\tau)$ of equation
(\ref{Gamma_properties}) guarantees that the coefficients $S_K$ are
real numbers. If the autocorrelation function is purely real then
the spectral density coefficients obey the relation $S_K=S_{-K}$.
But if $\Gamma(\tau)$ is complex then, in general, $S_K\neq S_{-K}$
and we refer to this as an asymmetry in the spectral density.

As an example we consider classical microwaves of frequency $\omega$
with magnetic flux of the form
\begin{eqnarray} \label{class_phi}
\phi(t) = \phi_1\sin(\omega t).
\end{eqnarray}
In this case the electron intensity at the point $x=0$ on the screen
is
\begin{eqnarray} \label{I_classical}
I_{\rm cl}(t)=1+\cos[e\phi_1\sin(\omega t)].
\end{eqnarray}
Therefore the autocorrelation function is
\begin{eqnarray} \label{G_cl}
\Gamma_{\rm cl}(\tau)=\left[1 + J_{0}(e \phi_1)\right]^2 +
2\sum_{K=1}^{\infty}\left[J_{2K}(e \phi_1) \right]^2 \cos(2K\omega\tau),
\end{eqnarray}
where the $J_{n}(z)$ are Bessel functions \cite{tables}, and the
spectral density coefficients are
\begin{eqnarray}
S_0=[1 + J_{0}(e \phi_1)]^2, \;\;\;\;\;\; S_K=[J_{2K}(e \phi_1)]^2.
\end{eqnarray}

%======================== FIGURE 5
\begin{figure} [ht]
\centering \scalebox{0.5} {\includegraphics{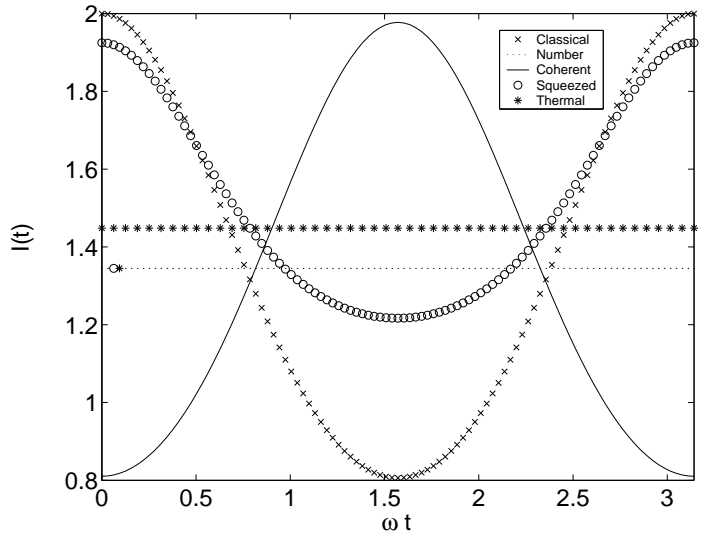}}\caption{$I(0,t)$ for the
electrons as a function of time, $\omega t$, for $\omega=10^{-4}$,
corresponding to irradiation with number states (broken line), coherent states
(solid line), squeezed states (line of circles), and thermal states (line of
stars). We have chosen $\langle N \rangle = 17$, in all four cases, and
$r=4.2$ for the squeezed states. For comparison, we also show the electron
intensity of equation (\ref{I_classical}) corresponding to classical
microwaves (line of crosses), for $\phi_1=(2\langle N\rangle)^{1/2}$.}
\end{figure}

%======================== FIGURE 6
\begin{figure} [ht]
\centering \scalebox{0.5} {\includegraphics{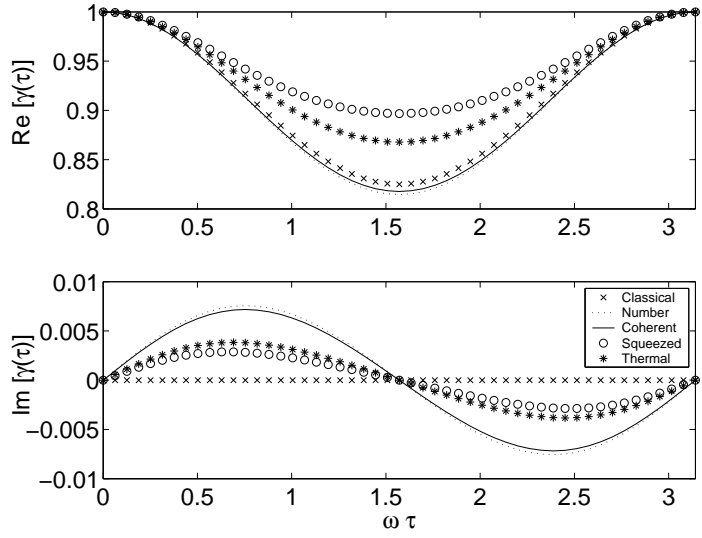}}\caption{Real and
imaginary parts of $\gamma(\tau)$ of equation (\ref{normalG}) for the
electrons as a function of time, $\omega \tau$, for $\omega=10^{-4}$,
corresponding to irradiation with number states (broken line), coherent states
(solid line), squeezed states (line of circles), and thermal states (line of
stars). The parameters are $\langle N \rangle = 17$, $r=4.2$, and for the case
of classical microwaves (line of crosses) we have $\phi_1=(2\langle N\rangle)
^{1/2}$.}
\end{figure}

\subsection{Autocorrelation function of the electron intensity in the case of nonclassical microwaves}
In this case the electron intensity $\hat{I}(t) = 1 +
\cos[e\hat{\phi}(t)]$ is an operator. Consequently the
autocorrelation function $\Gamma(\tau)$ of equation (\ref{autocorr})
is in this case defined as
\begin{eqnarray} \label{quantum_autocorr}
\Gamma(\tau) = \lim_{T\rightarrow \infty} \frac{1}{2T} \int_{-T}^{T} \Tr[\rho
\hat{I}^{\dag}(t)\hat{I}(t+\tau)] dt.
\end{eqnarray}

The values of $\Gamma(\tau)$ for the electric charges have been
derived for irradiation with various nonclassical microwave states
in \cite{Vourdas_2001}. In the following we present numerical
results, which illustrate the electron correlation properties, and
allow for a comparison between the effects of classical and
nonclassical microwaves.

%======================== FIGURE 7
\begin{figure} [ht]
\centering \scalebox{0.5} {\includegraphics{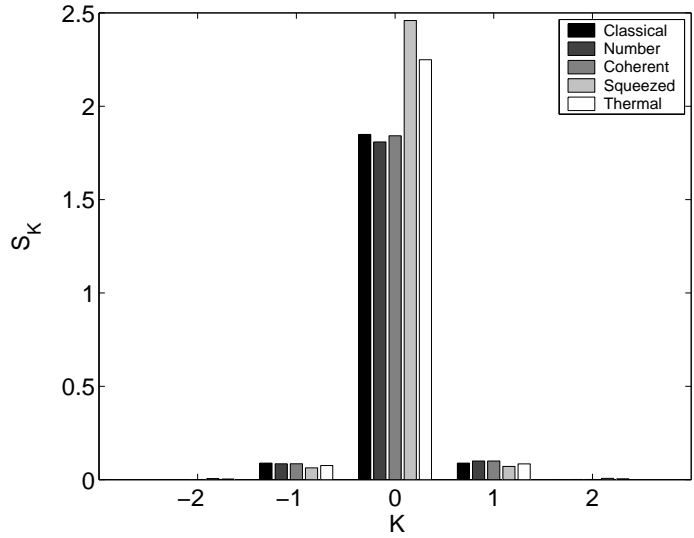}}\caption{Spectral density
coefficients $S_K$ of equation (\ref{psd}) for the electrons as a function of
$K$, corresponding to irradiation with classical microwaves (first column from
the left), and nonclassical microwaves in number states (second column),
coherent states (third column), squeezed states (fourth column), and thermal
states (fifth column). The parameters are $\langle N \rangle = 17$, $r=4.2$,
and $\phi_1=(2\langle N\rangle) ^{1/2}$.}
\end{figure}

\subsection{Numerical results}
In order to make the comparison meaningful, in the numerical
calculations (figures 5-7) the number of photons in the number
states is equal to the average number of photons in the coherent,
squeezed, and thermal states; we have chosen
\begin{eqnarray}
N = \langle N \rangle _{\rm coh} = \langle N \rangle _{\rm sq} = \langle N
\rangle _{\rm th} = 17.
\end{eqnarray}
For comparison with the case of classical microwaves we have chosen
the amplitude of the classical magnetic flux to be $\phi_1 =
(2\langle N\rangle) ^{1/2}$. The frequency of the microwaves is
$\omega = 10^{-4}$ in units where $k_B=\hbar =c =1$. The squeezing
parameter is $r=4.2$ and the other parameters are $\xi = 1$,
$\arg{A}=0$, $\varphi = 0$.

In figure 5 we show the electron intensity $I(t)$ for $x=0$ as a
function of $\omega t$ for irradiation with number states (broken
line), coherent states (solid line), squeezed states (line of
circles), and thermal states (line of stars). For comparison, we
also show the electron intensity of equation (\ref{I_classical})
corresponding to classical microwaves (line of crosses).

In figure 6 we show the real and imaginary parts of $\gamma(\tau)$
of equation (\ref{normalG}) for the electrons as a function of
$\omega \tau$ corresponding to irradiation with number states
(broken line), coherent states (solid line), squeezed states (line
of circles), and thermal states (line of stars); we have also
included the results in the case of classical microwaves (line of
crosses) for comparison. It is seen that different states of the
electromagnetic field lead to different electron correlation
properties. The imaginary part of the normalized autocorrelation
function vanishes only for irradiation with classical microwaves,
but it is nonzero for the four cases of nonclassical microwaves.

In figure 7 we plot the spectral density coefficients $S_K$ of
equation (\ref{psd}) for the electrons as a function of $K$,
corresponding to irradiation with classical microwaves (first column
from the left), and nonclassical microwaves in number states (second
column), coherent states (third column), squeezed states (fourth
column), and thermal states (fifth column).

%==============================================================================================================%
%----------------------------------- *** SECTION 7 *** --------------------------------------------------------7
\section{Entangled currents in distant electron interference experiments induced by entangled photons}
In this section we consider two electron interference devices that
are far from each other \cite{TCV}. A photon source irradiates the
two experiments with correlated two-mode nonclassical microwaves.
Each microwave field mode is coupled to one of the two experiments.
The experiment is depicted in figure 8. It will be shown that the
photon correlations are transferred to the electron interference
experiments, which become correlated. The nature of their
correlation depends on whether the external photons are separable
(classically correlated) or entangled (quantum mechanically
correlated).

%======================== FIGURE 8
\begin{figure}[ht]
\begin{center}
\scalebox{0.65}{\includegraphics{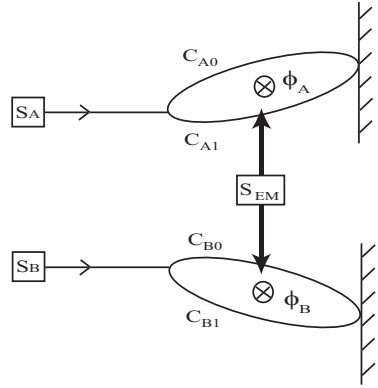}}
\end{center}
\caption{Two electron interference experiments which are far from each other
are irradiated with nonclassical electromagnetic fields. The two
electromagnetic fields in the two experiments are produced by the source
$S_{\rm EM}$ and are correlated.}
\end{figure}

Let $\rho$ be the density operator describing the two-mode
nonclassical electromagnetic field. The first mode of frequency
$\omega _1$ interacts with electrons in experiment ${\bf A}$ and its
density matrix is given by $\rho_{\rm A} = {\rm Tr}_{\rm B} (\rho)$.
Similarly the second mode of frequency $\omega _2$ interacts with
electrons in experiment ${\bf B}$ and its density matrix is
$\rho_{\rm B}={\rm Tr}_{\rm A} (\rho)$. The density matrix $\rho$
can be factorizable (i.e., the field modes are independent of each
other), separable (the field modes are classically correlated), or
entangled (the field modes are quantum mechanically correlated). The
difference between these cases has been discussed in section 2.

\subsection{Correlations of the electron intensity distributions}
The nonclassical magnetic flux $\hat{\phi}_{\rm A}$ that influences
the electron interference in \textbf{A} gives rise to the phase
factor operator $\exp(ie \hat{\phi}_{\rm A})$. This phase factor
induces the electron intensity distribution $I_{\rm A}(x_{\rm A})$,
which is given by
\begin{eqnarray} \label{I_quantum_A}
\fl I_{\rm A}(x_{\rm A})= \mbox{Tr}\{\rho_{\rm A}[1+\cos(x_{\rm
A}-e\hat\phi_{\rm A})]\} = 1 + |\tilde{W}(\lambda_{\rm A})|\cos\{x_{\rm A} -
\arg[ \tilde{W} (\lambda_ {\rm A})]\}
\end{eqnarray}
where $\lambda_{\rm A}=iq\exp(i\omega_1 t)$. Similarly in experiment
\textbf{B}, which is influenced by a nonclassical magnetic flux
$\hat{\phi}_{\rm B}$, one obtains the intensity
\begin{eqnarray} \label{I_quantum_B}
\fl I_{\rm B}(x_{\rm B})= \mbox{Tr}\{\rho_{\rm B}[1+\cos(x_{\rm
B}-e\hat\phi_{\rm B})]\} = 1 + |\tilde{W}(\lambda_{\rm B})|\cos\{x_{\rm B} -
\arg[ \tilde{W} (\lambda_ {\rm B})]\}
\end{eqnarray}
where $\lambda_{\rm B}=iq\exp(i\omega_2 t)$.

The electron intensity on the interference screen of experiment
\textbf{A} (or \textbf{B}) is calculated by tracing the intensity
operator with respect to the density matrix that describes the
corresponding electromagnetic field mode ($\rho_{\rm A}$ or
$\rho_{\rm B}$). The results, $I_{\rm A}(x_{\rm A})$ and $I_{\rm
B}(x_{\rm B})$, are proportional to the probability of detecting an
electron at a point $x_{\rm A}$ in \textbf{A} or a point $x_{\rm B}$
in \textbf{B}. It is also possible to define the joint electron
intensity $I(x_{\rm A},x_{\rm B})$, which is related to the
probability of a simultaneous detection of electrons at $x_{\rm A}$
and $x_{\rm B}$. This joint intensity is controlled by the full
density matrix $\rho$ for the two-mode electromagnetic field, that
is,
\begin{eqnarray} \label{I_quantum_AB}
I(x_{\rm A},x_{\rm B})= \mbox{Tr}\{\rho [1+\cos(x_{\rm A}-e\hat\phi_{\rm A})]
[1+\cos(x_{\rm B}-e\hat\phi_{\rm B})]\}.
\end{eqnarray}

If the two-mode electromagnetic field is factorizable ($\rho =
\rho_{\rm A} \otimes \rho_{\rm B}$) then the joint electron
intensity $I(x_{\rm A},x_{\rm B})$ is simply equal to the product
$I_{\rm A}(x_{\rm A})I_{\rm B}(x_{\rm B})$ of the independent
intensities. However, if the two field modes are classically or
quantum mechanically correlated then this is not true, in general.
In order to quantify this we can define the ratio
\begin{eqnarray} \label{ratio}
R =\frac { I(x_{\rm A} ,x_{\rm B} )} {I_{\rm A} (x_{\rm A} ) I_{\rm B} (x_{\rm
B})}
\end{eqnarray}
which is equal to one only for independent electron intensities. In
other words, whenever $R$ takes values not equal to one, the
electron intensity in experiment \textbf{A} is correlated to the
electron intensity in experiment \textbf{B}. In what follows we
consider particular examples that illustrate the effect.

\subsection{Examples and numerical results}
As an example we consider two-mode separable and entangled
microwaves in number states\cite{TCV}. We compare and contrast the
effects of the separable state
\begin{eqnarray} \label{num_rho_sep}
\rho_{\rm sep}=\frac{1}{2}(|00\rangle \langle 00|+ |11\rangle \langle 11|)
\end{eqnarray}
and the entangled state $ 2^{-1/2}(|00\rangle + |11\rangle)$ with
density matrix
\begin{eqnarray} \label{num_rho_ent}
\rho_{\rm ent} = \rho_{\rm sep} + \frac{1}{2}(|00 \rangle \langle 11| + | 11
\rangle \langle 00|).
\end{eqnarray}
It is noted that the $\rho_{\rm sep}$ and the $\rho_{\rm ent}$
differ only in the above nondiagonal elements. The reduced density
operators that describe the electromagnetic field in \textbf{A} and
\textbf{B} are, in both cases,
\begin{eqnarray} \label{num_reduced_rho}
\rho_{\rm sep,A}= \rho_{\rm ent,A} = \rho_{\rm sep,B} = \rho_{\rm ent,B} =
\frac{1}{2}(|0 \rangle \langle 0|+ |1\rangle \langle 1|).
\end{eqnarray}

%======================== FIGURE 9
\begin{figure}[htbp]
\begin{center}
\scalebox{0.5}{\includegraphics{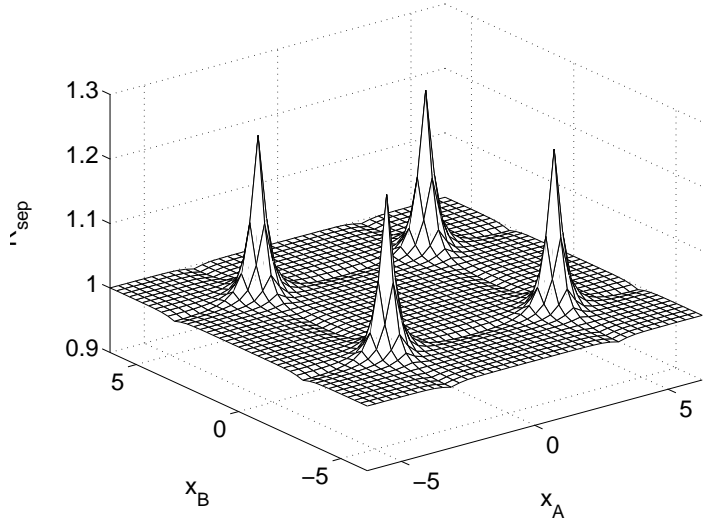}}
\end{center}
\caption{$R_{\rm sep}$ of equation (\ref{R_sep_num}) as a function of $x_{\rm
A},x_{\rm B} \in [-2\pi,2\pi]$. Here $\min(R_{\rm sep})=1.0001$ and
$\max(R_{\rm sep})=1.2471$. The frequencies are $\omega_1=1.2\times 10^{-4}$
and $\omega_2=10^{-4}$.}
\end{figure}

The ratio of equation (\ref{ratio}) corresponding to the separable
state $\rho_{\rm sep}$ is given by
\begin{eqnarray} \label{R_sep_num}
R_{\rm sep}(x_{\rm A},x_{\rm B})= \frac{1+\alpha(\cos x_{\rm A}+\cos x_{\rm
B}) +\gamma \cos x_{\rm A} \cos x_{\rm B}} {(1+\alpha\cos x_{\rm
A})(1+\alpha\cos x_{\rm B})}
\end{eqnarray}
where
\begin{eqnarray} \label{parameters_num}
\alpha=\frac{2-q^2}{2} \exp \left(-\frac{q^2}{2}\right), \;\;\;\;\;
\gamma=\frac{1}{2}\exp(-q^2)[1+(1-q^2)^2].
\end{eqnarray}
It can easily be shown that
\begin{eqnarray} \label{R_inequality}
\frac{1+2\alpha+\gamma}{(1+\alpha)^2} \leq R_{\rm sep}(x_{\rm A},x_{\rm B})
\leq \frac{1-2\alpha+\gamma}{(1-\alpha)^2}
\end{eqnarray}
which in our example leads to $\min (R_{\rm sep}) = 1.0001$ and
$\max (R_{\rm sep}) = 1.2471$. In figure 9 we plot the $R_{\rm
sep}(x_{\rm A},x_{\rm B})$ against screen positions $x_{\rm A}$ and
$x_{\rm B}$ for microwave frequencies $\omega_1=1.2\times 10^{-4}$
and $\omega_2=10^{-4}$.

%======================== FIGURE 10
\begin{figure}[htbp]
\begin{center}
\scalebox{0.5}{\includegraphics{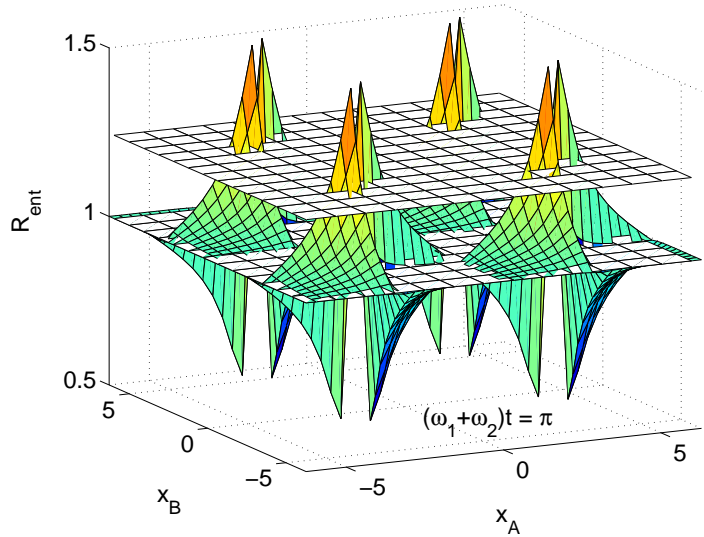}}
\end{center}
\caption{$R_{\rm ent}$ of equation (\ref{R_ent_num}) as a function of $x_{\rm
A},x_{\rm B} \in [-2\pi,2\pi]$ for $(\omega_1+\omega_2)t=\pi$. The top and
bottom plateaus show the $\max(R_{\rm sep})=1.2471$ and $\min(R_{\rm
sep})=1.0001$, respectively. The frequencies are $\omega_1=1.2\times 10^{-4}$
and $\omega_2=10^{-4}$.}
\end{figure}

In the case of the entangled state $\rho_{\rm ent}$ of equation
(\ref{num_rho_ent}) the ratio is
\begin{eqnarray} \label{R_ent_num}
\fl R_{\rm ent}(x_{\rm A},x_{\rm B},t)=R_{\rm sep}(x_{\rm A},x_{\rm B})
-q^2\exp(-q^2) \frac{\sin x_{\rm A} \sin x_{\rm B} \cos[(\omega_1+\omega_2)t]}
{(1+\alpha\cos x_{\rm A})(1+\alpha\cos x_{\rm B})}.
\end{eqnarray}
The $R_{\rm ent}$ oscillates in time around the $R_{\rm sep}$, with
frequency $\omega_1+\omega_2$, and exceeds periodically the bounds
of the inequality for $R_{\rm sep}$ in (\ref{R_inequality}). In
figure 10 we plot the $R_{\rm ent}(x_{\rm A},x_{\rm B})$ against
screen positions $x_{\rm A}$ and $x_{\rm B}$ for
$(\omega_1+\omega_2)t=\pi$ and the same microwave frequencies as in
the previous figure. The two horizontal surfaces represent the $\min
(R_{\rm sep})$ (bottom plateaux) and the $\max (R_{\rm sep})$ (top
plateaux). In figure 11 we compare the $R_{\rm sep}$ (line of
circles) and the $R_{\rm ent}$ (solid line) against time
$(\omega_1+\omega_2)t$ for fixed screen positions $x_{\rm A}=0.9\pi$
and $x_{\rm B}=1.025\pi$.

%======================== FIGURE 11
\begin{figure}[htbp]
\begin{center}
\scalebox{0.5}{\includegraphics{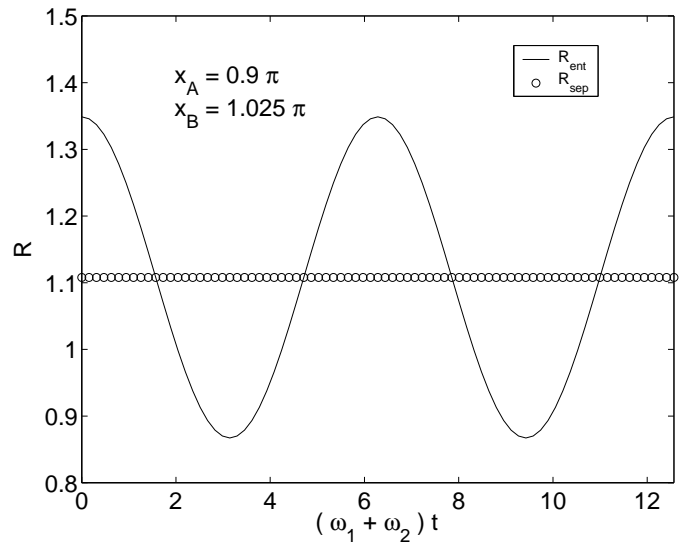}}
\end{center}
\caption{Comparison of $R_{\rm ent}$ (solid line) and $R_{\rm sep}$ (line of
circles) of equations (\ref{R_ent_num}) and (\ref{R_sep_num}), respectively,
for $x_{\rm A}=0.9\pi$ and $x_{\rm B}=1.025\pi$ as a function of dimensionless
time. The frequencies are $\omega_1=1.2\times 10^{-4}$ and
$\omega_2=10^{-4}$.}
\end{figure}

%==============================================================================================================%
%----------------------------------- *** SECTION 8 *** --------------------------------------------------------8
\section{Interaction of mesoscopic SQUID rings with nonclassical electromagnetic fields}
In this section we investigate application of the above ideas in the
context of mesoscopic superconducting quantum interference device
(SQUID) rings. In the first instance we introduce mesoscopic SQUID
rings and describe how they interact with nonclassical
electromagnetic fields \cite{Jos_Vourdas94, Jos_Vourdas96}. In this
case the Josephson currents are quantum mechanical operators, whose
expectation values with respect to the density operator of the
external photons, yield the observed currents. Subsequently, we
apply the general concept described in the previous section to the
case of two distant SQUID rings, each of which is coupled to a
single mode of a two-mode nonclassical electromagnetic field
\cite{TCV_SQUID}. It is shown that the photon correlations are
transferred to the Josephson currents in the distant superconducting
devices.

\subsection{Mesoscopic SQUID ring}
Consider a superconducting ring of mesoscopic area $\xi \leq
10^{-8}{\rm cm}^2$, which is interrupted by a Josephson junction
(weak link), as shown in the figure below. In this case the
capacitance $C$ across the Josephson junction is very small and at
low temperatures $T < 0.1K$ the behaviour of the SQUID is
nonclassical \cite{SQUID_2} in the sense that the Coulomb charging
energy for a Cooper pair of charge $2e$,
\begin{eqnarray}
E_C=\frac{(2e)^2}{C},
\end{eqnarray}
becomes a significant parameter.

%======================== FIGURE 12
\begin{figure} [ht]
\begin{center}
\scalebox{0.20}{\includegraphics{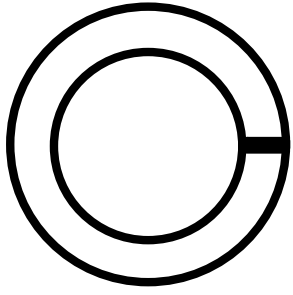}}
\end{center}
\caption{Mesoscopic SQUID ring: superconducting ring of mesoscopic
dimensions is interrupted by a Josephson junction.}
\end{figure}

Under these conditions the charge $\hat Q = - i (2e)
\partial_{\delta}$ through the junction and the phase difference
across the junction $\hat \delta$ are conjugate operators, which
obey the commutation relation $[\hat\delta,\hat Q] = i2e$. In this
case the Josephson current also becomes an operator which is a
sinusoidal function of the phase difference across the junction,
\begin{eqnarray} \label{quantum_current}
\hat I = I_{\rm c} \sin\hat\delta,
\end{eqnarray}
where $I_{\rm c}$ is the critical current.

\subsection{Interaction with classical microwaves}
We consider a mesoscopic SQUID ring interacting with a monochromatic
electromagnetic field. The  magnetic flux $\phi(t)$ is threading the
SQUID ring and the phase difference across the junction is $\delta =
2e \phi(t)$.

In the classical case the Josephson current $I$ and the phase
difference $\delta$, are classical numbers. Therefore
for a magnetic flux with a linear and a sinusoidal component,
\begin{eqnarray} \label{phi_class_intoSQUID}
\phi(t) = \phi_0 + V_{1} t + u \sin(\omega_1 t),
\end{eqnarray}
we get the current
\begin{eqnarray} \label{I_class_1}
I = I_1 \sin[2e\phi(t)] = I_1 \sin[2e\phi_0 + 2e V_{1} t + 2e u \sin(\omega_1
t)].
\end{eqnarray}

Using the well-known identity
\begin{eqnarray} \label{identity_exp}
\exp(iu\sin z)=\sum_{n=-\infty}^{\infty} J_{n}(u) \exp(i n z),
\end{eqnarray}
we can easily show that
the current can be expanded as
\begin{eqnarray} \label{I_class_2}
I = I_1 \sum_{n=-\infty}^{\infty} J_{n}(2e u) \sin[(2e V_{1} + n\omega_1) t +
2e\phi_0].
\end{eqnarray}
Calculating the time-averaged value $I_{\rm dc}$ of the current $I$
we see that when
\begin{eqnarray} \label{shapiro_steps_class}
2eV_0 = N\omega
\end{eqnarray}
where $N$ is an integer, we get
\begin{eqnarray} \label{I_dc_class}
I_{\rm dc}=I_1 J_{-N}(2e u) \sin(2e\phi_0),
\end{eqnarray}
otherwise the $I_{\rm dc}$ vanishes. These integral values of the
voltage are usually referred to as Shapiro steps.

\subsection{Interaction with nonclassical microwaves}
We now study the effect of nonclassical microwaves on the Josephson
current of a mesoscopic SQUID ring operating at low temperatures.We
use the external field approximation, and ignore the back-reaction.
This is a good approximation when the external electromagnetic
fields are much stronger than the fields induced by the currents
circulating the mesoscopic devices.

We consider the irradiation of a mesoscopic SQUID ring with
monochromatic nonclassical microwaves of frequency $\omega_1$. In
addition to that the ring is threaded by the classical flux $\phi
_0+V_1 t$ and the total flux is $\hat\Phi(t)=\phi
_0+V_1t+\hat\phi(t)$. Therefore the quantum current is in this case
given by
\begin{eqnarray}
\hat I_{\rm A} &=& I_1 \sin \left\{ 2e\phi_0+2e V_1 t+ q'
\left[\exp(i\omega_1 t)\hat{a}^\dagger + \exp(-i\omega_1 t)\hat{a}\right]
\right\} \nonumber \\
&=& I_1 \Im \left\{\exp[i(\omega_A t+2e\phi_0)] D[iq'\exp(i\omega_1 t)]\right\}
\end{eqnarray}
where
\begin{eqnarray} \label{q_definition_2}
\omega_A=2e V_1, \;\;\;\;\; q'=\sqrt{2}e \xi.
\end{eqnarray}
It is noted that the scaled electric charge $q'$ has twice the value
of $q$ of equation (\ref{q_definition}), because in this case we
have pairs of electrons. The experimentally measured current is
calculated by tracing with respect to the density operator
$\rho_{\rm A}$ for the external electromagnetic fields, that is,
\begin{eqnarray} \label{I_A_SQUID}
\langle I_{\rm A}\rangle \equiv \Tr (\rho_{\rm A} \hat I_{\rm A}) = I_1
\Im[\exp (i\omega_{\rm A}t) \tilde{W}(\sigma_{\rm A})], \;\;\;\;\; \sigma_{\rm
A} = iq' \exp(i\omega_1 t).
\end{eqnarray}

As an example we consider microwaves in coherent states. For
comparison with the classical case of equation (\ref{I_dc_class}) we
take coherent states with $A=2^{-1/2}u$ and $\arg A=0$. In this case
we get Shapiro steps, as in the classical case, but the dc current
is now reduced by a small factor:
\begin{eqnarray} \label{I_dc_coh}
I_{\rm dc}^{\rm (coh)} =
\exp\left(- \frac{{q'}^{2}}{2} \right) I_{\rm dc}.
\end{eqnarray}

We also consider the case where the microwaves are in a squeezed
vacuum. The squeezed vacuum is a superposition of even number states
only. In this case \cite{Jos_Vourdas94} we get even Shapiro steps
only. A physical interpretation of this result is that the electrons
can only absorb an even number of photons (there are no odd number
states in this quantum state). Similar results can be proved for
even Schroedinger cats (${\cal N}(|A\rangle + |-A\rangle $), which
are superpositions of even number states also. We stress that in
this case the result is qualitatively different from the classical
result, in the sense that the odd Shapiro steps are absent.

\subsection{Entanglement of distant mesoscopic SQUID rings}
We consider two spatially separated mesoscopic SQUID rings, which we
refer to as A and B. They are irradiated with microwaves that are
described by a density operator $\rho$. The microwaves are produced
by the same source and are correlated. Photons of frequency
$\omega_1$ interact with device A;and photons of frequency
$\omega_2$ interact with device B. The SQUID rings A and B are also
threaded by a classical time-dependent magnetic fluxes that increase
linearly with time ($V_{\rm A}t$ and $V_{\rm B}t$, respectively).
The proposed experiment is illustrated in figure 13.

%======================== FIGURE 13
\begin{figure} [ht]
\begin{center}
\scalebox{0.35}{\includegraphics{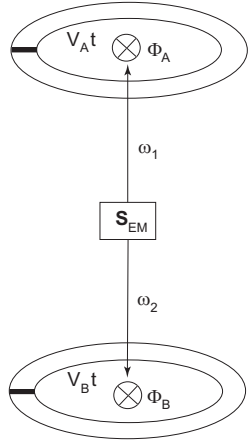}}
\end{center}
\caption{Two distant mesoscopic SQUID rings {\rm A} and {\rm B} are irradiated
with nonclassical microwaves of frequencies $\omega_1$ and $\omega_2$,
respectively. The microwaves are produced by the source ${\rm S}_{\rm EM}$ and
are correlated. Classical magnetic fluxes $V_{\rm A}t$ and $V_{\rm B}t$ are
also threading the two rings {\rm A} and {\rm B}, respectively.}
\end{figure}

The observed Josephson current in SQUID ring A or B is given by the
expectation value of the corresponding current operator,
\begin{eqnarray}
\langle \hat I_{\rm A}\rangle &=& I_{1} \mbox{Tr}(\rho_{\rm A}\sin \hat
\delta_{\rm A}), \;\;\;\;\; \hat \delta_{\rm A}= 2eV_{\rm A}t + 2e
\hat\phi_{\rm A}(t),
\label{I_A_c} \\
\langle \hat I_{\rm B}\rangle &=& I_{2} \mbox{Tr}(\rho_{\rm B}\sin \hat
\delta_{\rm B}), \;\;\;\;\; \hat \delta_{\rm B}= 2eV_{\rm B}t + 2e
\hat\phi_{\rm B}(t), \label{I_B_c}
\end{eqnarray}
where
\begin{eqnarray}
\hat \phi_{\rm A}(t)=\frac{\xi} {\sqrt{2}} \left[\exp(i\omega_1
t)\hat{a_1}^\dagger + \exp(-i\omega_1 t)\hat{a_1} \right], \\
\hat \phi_{\rm B}(t)=\frac{\xi} {\sqrt{2}} \left[\exp(i\omega_2 t)
\hat{a_2}^\dagger + \exp(-i\omega_2 t)\hat{a_2}\right],
\end{eqnarray}
in accordance with the formalism developed in section 3 and assuming
that both rings have the same area $\xi$. The $\langle \hat I_{\rm
A}\rangle$ has been written in terms of the Weyl function $\tilde
W(\sigma_{\rm A})$ in equation (\ref{I_A_SQUID}); and similarly for
B.

The expectation value of the product of the two current operators is
given by:
\begin{equation}\label{I_AB}
\langle\hat I_{\rm A}\hat I_{\rm B}\rangle = I_{1}I_{2} \mbox{Tr}(\rho
\sin\hat\delta_{\rm A}\sin\hat\delta_{\rm B}).
\end{equation}
The correlations between the observed electron currents can be
quantified by defining the ratio
\begin{equation}\label{ratio_c}
R^{\rm (c)} = \frac{\langle\hat I_{\rm A} \hat I_{\rm B}\rangle} {\langle\hat
I_{\rm A}\rangle \langle\hat I_{\rm B}\rangle},
\end{equation}
where the superscript $(c)$ indicates that this quantity corresponds
to currents. For factorizable density matrices $\rho_{\rm
fact}=\rho_{A}\otimes\rho_B$ we easily obtain the ratio $R_{\rm
fact}^{\rm (c)}=1$, identically. For separable density matrices
$\rho_{\rm sep}$ of equation (\ref{rho_sep}) we get
\begin{eqnarray}
R_{\rm sep}^{\rm (c)} = \frac{\sum_i p_i \langle \hat I_{{\rm
A}i}\rangle\langle \hat I_{{\rm B}i}\rangle} {(\sum_k p_k \langle \hat I_{{\rm
A}k}\rangle)(\sum_l p_l \langle \hat I_{{\rm B}l}\rangle)}.
\end{eqnarray}

We also calculate the higher moments of the currents
\begin{eqnarray}
\langle {\hat I}_{\rm A}^2 \rangle &=& I_{1}^2 \mbox{Tr}[\rho_{\rm A}(\sin
\hat\sigma_{\rm A})^2], \label{I_A_2} \\
\langle {\hat I}_{\rm B}^2\rangle &=& I_{2}^2 \mbox{Tr}[\rho_{\rm B}(\sin \hat
\sigma_{\rm B})^2], \label{I_B_2} \\
\langle {\hat I}_{\rm A}^2{\hat I}_{\rm B}^2 \rangle &=& I_{1}^2 I_{2}^2
\mbox{Tr}[\rho (\sin {\hat\sigma}_{\rm A})^2(\sin {\hat \sigma} _{\rm B})^2].
\label{I_AB_2}
\end{eqnarray}
The expectation value $\langle {\hat I}_{\rm A}^M {\hat I}_{\rm B}^N
\rangle$ quantifies the quantum statistics of the electron pairs
tunneling the junctions in the two SQUID rings. Consequently the
ratio
\begin{eqnarray}\label{ratio_R2}
R^{\rm (c2)} = \frac{\langle {\hat I}_{\rm A}^2 {\hat I}_{\rm B}^2\rangle}
{\langle{\hat I}_{\rm A}^2\rangle \langle{\hat I}_{\rm B}^2\rangle}
\end{eqnarray}
is a measure of the photon-induced correlations of the quantum
statistics of the tunneling electrons. For factorizable density
matrices we easily see that $R^{\rm (c2)}_{\rm fact}=1$. For
separable density matrices we get
\begin{eqnarray}
R^{\rm (c2)}_{\rm sep} = \frac{\sum_i p_i \langle {\hat I}_{{\rm
A}i}^2\rangle\langle {\hat I}_{{\rm B}i}^2\rangle} {(\sum_k p_k \langle {\hat
I}_{{\rm A}k}^2\rangle )(\sum_l p_l \langle {\hat I}_{{\rm B}l}^2\rangle)}.
\end{eqnarray}

\subsection{Examples and numerical results}
We present examples in which we compare and contrast the influence
of a classically correlated two-mode microwave state with a quantum
mechanically correlated one, on the Josephson currents.  The
two-mode microwaves are in both number and coherent states.

\subsubsection{Number states}
Firstly we consider the separable density operator $\rho_{\rm sep}$
of equation (\ref{rho_sep_num}) and the entangled density operator
$\rho_{\rm ent}$ of equation (\ref{rho_ent_num}) for number states.

For the $\rho_{\rm sep}$ of equation (\ref{rho_sep_num}) we
calculate the currents in A and B:
\begin{eqnarray}
\langle\hat I_{\rm A}\rangle &=& I_{1} C_0 \sin(\omega_{\rm A}t), \label{I_A_num}\\
\langle\hat I_{\rm B}\rangle &=& I_{2} C_0 \sin(\omega_{\rm B}t), \label{I_B_num}\\
C_0&=&\frac{1}{2}\exp\left(-\frac{q^{'2}}{2}\right)[L_{N_1}(q^{'2})+L_{N_2}(q^{'2})],\label{C}
\end{eqnarray}
where the $L_{n}^{\alpha}(x)$ are Laguerre polynomials
\cite{tables}. It is noted that in this case the currents
$\langle\hat I_{\rm A}\rangle, \langle\hat I_{\rm B}\rangle$ are
independent of the microwave frequencies $\omega_1,\omega_2$. The
second moments of the currents in A and B, defined by equations
(\ref{I_A_2}) and (\ref{I_B_2}), respectively, have also been
calculated:
\begin{eqnarray}
\langle \hat I_{\rm A}^2 \rangle &=&  \frac{I_{1}^2}{2}
[1 - C_1 \cos(2\omega_{\rm A}t) ], \label{I_A_squared} \\
\langle \hat I_{\rm B}^2 \rangle &=&  \frac{I_{2}^2}{2}
[1 - C_1 \cos(2\omega_{\rm B}t)], \label{I_B_squared} \\
C_1 &=& \frac{1}{2}\exp(-2q^{'2})[L_{N_1}(4q^{'2})+L_{N_2}(4q^{'2})].
\end{eqnarray}
The expectation value of the product of the two currents is
\begin{eqnarray} \label{I_c_sep}
\langle{\hat I}_{\rm A} {\hat I}_{\rm B}\rangle_{\rm sep}&=& I_{1}I_{2} C_2
\sin(\omega_{\rm A}t)\sin(\omega_{\rm B}t), \\
C_2 &=& \exp(-q^{'2}) L_{N_1}(q^{'2})L_{N_2}(q^{'2}).
\end{eqnarray}
Consequently the ratio $R^{(c)}$ of equation (\ref{ratio_c}) is
\begin{eqnarray}\label{R_sep_c}
R_{\rm sep}^{(c)}=\frac{C_2}{C_{0}^{2}}=
\frac{4L_{N_1}(q^{'2})L_{N_2}(q^{'2})} {[L_{N_1}(q^{'2})+L_{N_2}(q^{'2})]^2}.
\end{eqnarray}
In this example the $R_{\rm sep}^{(c)}$ is time-independent; it
depends only on the number of photons $N_1,N_2$ in the two-mode
microwave field.

%======================== FIGURE 14
\begin{figure}[htbp]
\begin{center}
\scalebox{0.5}{\includegraphics{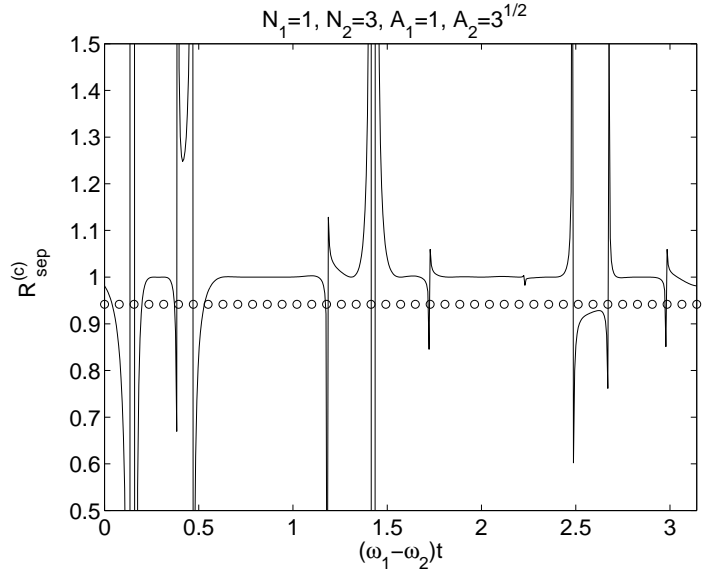}}
\end{center}
\caption{Comparison of $R^{\rm (c)}_{\rm sep}$ for the separable number state
of equation (\ref{rho_sep_num}) (line of circles) and $R^{\rm (c)}_{\rm sep}$
for the separable coherent state of equation (\ref{rho_sep_coherent}) (solid
line) for $N_1=1,N_2=3$ and $A_1=1,A_2=3^{1/2}$ as a function of
$(\omega_1-\omega_2)t$, where $\omega_1=1.2\times 10^{-4}$ and
$\omega_2=10^{-4}$.}
\end{figure}

For the $\rho_{\rm ent}$ of equation (\ref{rho_ent_num}) the
$\langle\hat I_{\rm A}\rangle,\langle\hat I_{\rm B}\rangle$ are the
same with those presented in equations (\ref{I_A_num}),
(\ref{I_B_num}); and the $\langle\hat I_{\rm A}^2\rangle,\langle\hat
I_{\rm B}^2\rangle$ are the same as in equations
(\ref{I_A_squared}), (\ref{I_B_squared}). However the $\langle{\hat
I}_{\rm A}{\hat I}_{\rm B}\rangle$ is in this case
\begin{equation}\label{I_c_ent}
\langle{\hat I}_{\rm A}{\hat I}_{\rm B}\rangle_{\rm ent} = \langle{\hat
I}_{\rm A}{\hat I}_{\rm B}\rangle_{\rm sep} + I_{\rm cross},
\end{equation}
where
\begin{eqnarray} \label{I_cross}
\fl I_{\rm cross} &=& -I_{1}I_{2}C_3 [\cos(\omega_{\rm A}t+\omega_{\rm
B}t)-(-1)^{N_1-N_2} \cos(\omega_{\rm A}t-\omega_{\rm B}t)] \cos(\Omega t), \\
\fl C_3 &=&\frac{1}{2} \exp(-q^{'2}) L_{N_1}^{N_2-N_1}(q^{'2})
L_{N_2}^{N_1-N_2}(q^{'2}).
\end{eqnarray}
The $I_{\rm cross}$ causes the $\langle{\hat I}_{\rm A}{\hat I}_{\rm
B}\rangle_{\rm ent}$ to oscillate in time around the $\langle{\hat I}_{\rm
A}{\hat I}_{\rm B}\rangle_{\rm sep}$ with frequency
\begin{equation}\label{Omega}
\Omega=(N_1-N_2)(\omega_1-\omega_2).
\end{equation}
We note that the term $I_{\rm cross}$ is induced by the nondiagonal
elements of $\rho_{\rm ent}$ of equation (\ref{rho_ent_num}), and
depends on the photon frequencies $\omega_1,\omega_2$. This term
quantifies the difference between the effect of separable and
entangled microwaves on the Josephson currents.

%======================== FIGURE 15
\begin{figure}[htbp]
\begin{center}
\scalebox{0.5}{\includegraphics{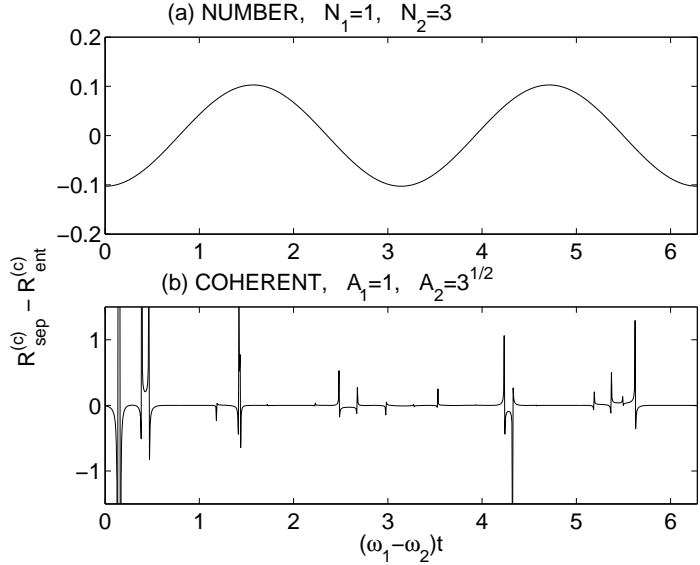}}
\end{center}
\caption{The difference $R^{\rm (c)}_{\rm sep} - R^{\rm (c)}_{\rm ent}$
corresponding to (a) the separable and entangled number states of equations
(\ref{rho_sep_num}), (\ref{rho_ent_num}); and (b) the separable and entangled
coherent states of equations (\ref{rho_sep_coherent}),
(\ref{rho_ent_coherent}), for $N_1=1,N_2=3$ and $A_1=1,A_2=3^{1/2}$ as a
function of $(\omega_1-\omega_2)t$, where $\omega_1=1.2\times 10^{-4}$ and
$\omega_2=10^{-4}$.}
\end{figure}

In the entangled case the ratio $R^{(c)}$ of equation
(\ref{ratio_c}) can be simplified in two distinct expressions
according to whether the difference $N_1-N_2$ is even or odd. In the
case $N_1-N_2=2k$, the ratio is
\begin{eqnarray}\label{R_ent_even}
R_{{\rm ent},2k}^{(c)}=R_{\rm sep}^{(c)}+\frac{4L_{N_1}^{-2k}
(q^{'2})L_{N_2}^{2k}(q^{'2})} {[L_{N_1}(q^{'2})+L_{N_2} (q^{'2})]^2}
\cos(\Omega t).
\end{eqnarray}
It is seen that the $R_{{\rm ent},2k}^{(c)}$ oscillates around the
$R_{\rm sep}^{(c)}$ with frequency $\Omega$ given by equation
(\ref{Omega}). If there is no detuning between the nonclassical
electromagnetic fields, i.e. $\omega_1=\omega_2$, then $R_{{\rm
ent},2k}^{(c)}$ is constant, although it is still $R_{\rm
ent}^{(c)}\neq R_{\rm sep}^{(c)}$. In the case $N_1-N_2=2k+1$ the
ratio is
\begin{eqnarray}\label{R_ent_odd}
R_{{\rm ent},2k+1}^{(c)}= R_{\rm sep}^{(c)}
-\frac{4L_{N_1}^{-2k-1}(q^{'2})L_{N_2}^{2k+1}(q^{'2})}
{[L_{N_1}(q^{'2})+L_{N_2}(q^{'2})]^2}\frac{\cos(\Omega t)} {\tan(\omega_{\rm
A}t)\tan(\omega_{\rm B}t)}.
\end{eqnarray}
In both cases the $R_{\rm ent}^{(c)}$ is time-dependent and it is a
function of the photon frequencies $\omega_1,\omega_2$, in contrast
to the case of $R_{\rm sep}^{(c)}$ (which is time-independent).

%======================== FIGURE 16
\begin{figure}[htbp]
\begin{center}
\scalebox{0.5}{\includegraphics{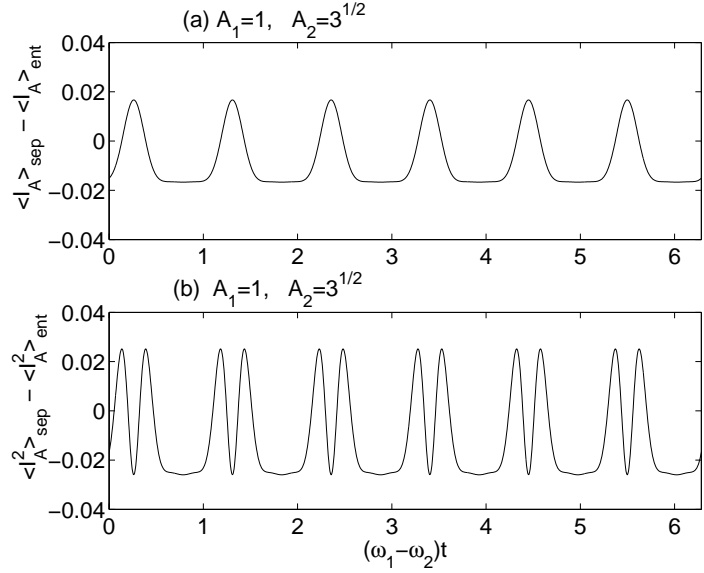}}
\end{center}
\caption{Difference of currents defined by equations (\ref{I_A_SQUID}) and
(\ref{I_A_2}) that are induced by separable and entangled photons in coherent
states (for SQUID ring A). (a) $\langle I_{\rm A}\rangle _{\rm sep} - \langle
I_{\rm A}\rangle _{\rm ent}$ and (b) $\langle I_{\rm A}^{2}\rangle _{\rm sep}
- \langle I_{\rm A}^{2} \rangle _{\rm ent}$ corresponding to irradiation with
separable and entangled coherent states of equations (\ref{rho_sep_coherent})
and (\ref{rho_ent_coherent}), for $A_1=1,A_2=3^{1/2}$ as a function of
$(\omega_1-\omega_2)t$, where $\omega_1=1.2\times 10^{-4}$ and
$\omega_2=10^{-4}$.}
\end{figure}

%======================== FIGURE 17
\begin{figure}[htbp]
\begin{center}
\scalebox{0.5}{\includegraphics{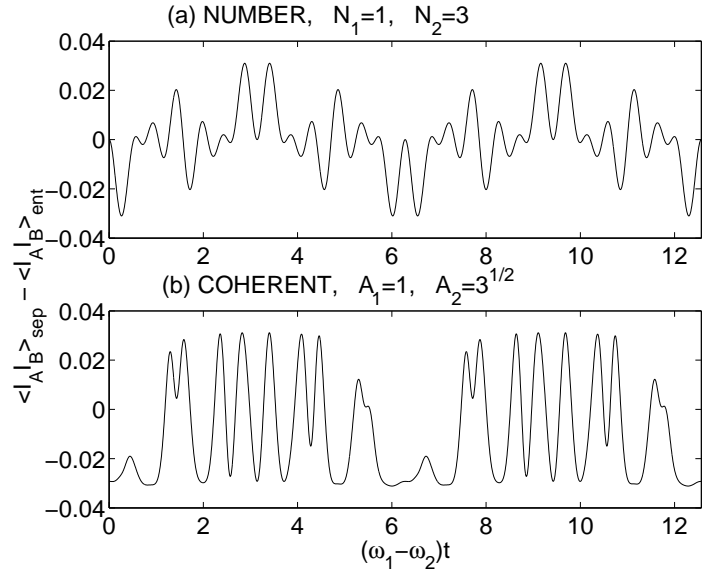}}
\end{center}
\caption{Difference of the product of currents $\langle I_{\rm A} I_{\rm B}
\rangle _{\rm sep} - \langle I_{\rm A} I_{\rm B} \rangle _{\rm ent}$, defined
by equation (\ref{I_AB}), that are induced by (a) the separable and entangled
number states of equations (\ref{rho_sep_num}), (\ref{rho_ent_num}); and (b)
the separable and entangled coherent states of equations
(\ref{rho_sep_coherent}), (\ref{rho_ent_coherent}), for $N_1=1,N_2=3$ and
$A_1=1,A_2=3^{1/2}$, respectively, as a function of $(\omega_1-\omega_2)t$,
where $\omega_1=1.2\times 10^{-4}$ and $\omega_2=10^{-4}$.}
\end{figure}

\subsubsection{Coherent states}
We consider the separable density operator $\rho_{\rm sep}$ of
equation (\ref{rho_sep_coherent}) and the entangled density operator
$\rho_{\rm ent}$ of equation (\ref{rho_ent_coherent}).

For the separable state of equation (\ref{rho_sep_coherent}) the currents in A
and B are
\begin{eqnarray} \label{I_A_sep}
\langle\hat I_{\rm A}\rangle_{\rm sep} &=& \frac{I_{1}}{2}
\exp(-\frac{q^{'2}}{2}) \{\sin[\omega_{\rm A}t
+2q'|A_1|\cos(\omega_1 t-\theta_{1})] \nonumber \\
&+& \sin[\omega_{\rm A}t+2q'|A_2|\cos(\omega_1 t-\theta_{2})]\},
\end{eqnarray}
\begin{eqnarray} \label{I_B_sep}
\langle\hat I_{\rm B}\rangle_{\rm sep} &=&
\frac{I_{2}}{2}\exp(-\frac{q^{'2}}{2})
\{\sin[\omega_{\rm B}t+2q'|A_1|\cos(\omega_2 t-\theta_{1})] \nonumber \\
&+& \sin[\omega_{\rm B}t+2q'|A_2|\cos(\omega_2 t-\theta_{2})]\},
\end{eqnarray}
where $\theta_1=\arg (A_1)$, and $\theta_2=\arg (A_2)$. The
expectation values of the product of the currents, and hence the
ratios $R_{\rm sep}^{(c)}$ and $R_{\rm sep}^{(c2)}$, have been
calculated numerically.

%======================== FIGURE 18
\begin{figure}[htbp]
\begin{center}
\scalebox{0.5}{\includegraphics{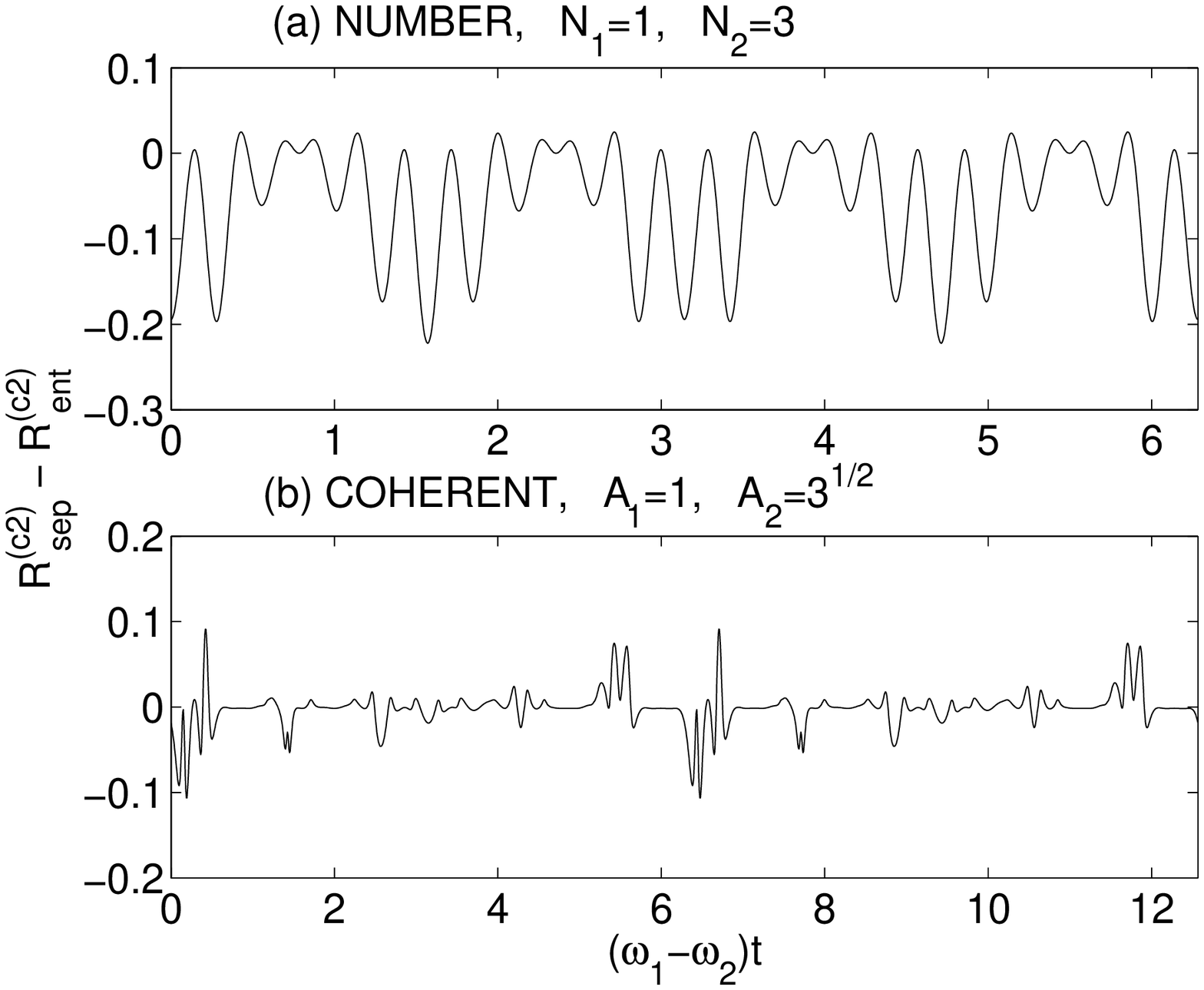}}
\end{center}
\caption{The difference $R^{\rm (c2)}_{\rm sep} - R^{\rm (c2)}_{\rm ent}$
corresponding to (a) the separable and entangled number states of equations
(\ref{rho_sep_num}), (\ref{rho_ent_num}); and (b) the separable and entangled
coherent states of equations (\ref{rho_sep_coherent}),
(\ref{rho_ent_coherent}), for $N_1=1,N_2=3$ and $A_1=1,A_2=3^{1/2}$ as a
function of $(\omega_1-\omega_2)t$, where $\omega_1=1.2\times 10^{-4}$ and
$\omega_2=10^{-4}$.}
\end{figure}

For the entangled state of equation (\ref{rho_ent_coherent}) the
current in A is
\begin{eqnarray} \label{I_A_ent}
\langle\hat I_{\rm A}\rangle_{\rm ent}= 2{\cal N}^2 \langle\hat I_{\rm
A}\rangle_{\rm sep}+ {\cal N}^2 E F_1 \exp\left(-\frac{q^{'2}}{2}\right)I_{1},
\end{eqnarray}
where
\begin{eqnarray}
E = \exp[-|A_1|^2 - |A_2|^2+2|A_1 A_2|\cos(\theta_{1}-\theta_{2})],
\end{eqnarray}
and
\begin{eqnarray}
\fl F_1 &=& \{ \exp[q|A_1|S_{A,1}(t)-q|A_2|S_{A,2}(t)]
+\exp[-q'|A_1|S_{A,1}(t) + q'|A_2|S_{A,2}(t)]\} \nonumber \\
\fl &\times & \sin[\omega_A t+q'|A_1|C_{A,1}(t)+q'|A_2|C_{A,2}(t)].
\end{eqnarray}
The terms entering the factor $F_1$ are trigonometric functions of
the form
\begin{eqnarray}
S_{A,1}&=&\sin(\omega_1 t-\theta_{1}), \;\;\;\;\;
S_{A,2}=\sin(\omega_1 t-\theta_{2}), \nonumber \\
C_{A,1}&=&\cos(\omega_1 t-\theta_{1}), \;\;\;\;\;
C_{A,2}=\cos(\omega_1t-\theta_{2}).
\end{eqnarray}
Similarly the current in SQUID ring B is
\begin{eqnarray}\label{I_B_ent}
\langle\hat I_{\rm B}\rangle_{\rm ent} =2{\cal N}^2 \langle\hat I_{\rm
B}\rangle_{\rm sep}+ {\cal N}^2 E F_2 \exp\left(-\frac{q^{'2}}{2}\right)I_{2},
\end{eqnarray}
where
\begin{eqnarray}
\fl F_2 &=&\{\exp[q|A_1|S_{B,1}(t)-q|A_2|S_{B,2}(t)]
+\exp[-q'|A_1|S_{B,1}(t)+q'|A_2|S_{B,2}(t)] \} \nonumber \\
\fl & \times &\sin[\omega_B t+q'|A_1|C_{B,1}(t)+q'|A_2|C_{B,2}(t)],
\end{eqnarray}
and
\begin{eqnarray}
S_{B,1}&=&\sin(\omega_2 t-\theta_{1}), \;\;\;\;\; S_{B,2}=\sin(\omega_2 t-\theta_{2}),\nonumber \\
C_{B,1}&=&\cos(\omega_2t-\theta_{1}), \;\;\;\;\;
C_{B,2}=\cos(\omega_2t-\theta_{2}).
\end{eqnarray}
The expectation values of the product of the currents, and hence the
ratios $R_{\rm ent}^{(c)}$ and $R_{\rm ent}^{(c2)}$, have been
calculated numerically.

\subsubsection{Numerical results}
In figures 14-18 we plot the results against dimensionless time
$(\omega_1-\omega_2)t$, where the photon frequencies are
$\omega_1=1.2\times 10^{-4}$ and $\omega_2=10^{-4}$. Other fixed
parameters are the number of photons in the number states:
$N_1=1,N_2=3$; and the average number of photons in the coherent
states: $A_1=1,A_2=3^{1/2}$ (we take these values so that the
microwaves in number and coherent states contain the same average
number of photons).

In figure 14 we present the $R^{\rm (c)}_{\rm sep}$ for the
separable number state of equation (\ref{rho_sep_num}) (line of
circles) and the $R^{\rm (c)}_{\rm sep}$ for the separable coherent
state of equation (\ref{rho_sep_coherent}) (solid line). It is seen
that separable photons in different quantum states induce different
correlations $R^{\rm (c)}$ between the Josephson currents in the
distant SQUID rings.

In figure 15 we show the difference $R^{\rm (c)}_{\rm sep} - R^{\rm
(c)}_{\rm ent}$ corresponding to (a) the separable and entangled
number states of equations (\ref{rho_sep_num}), (\ref{rho_ent_num});
and (b) the separable and entangled coherent states of equations
(\ref{rho_sep_coherent}), (\ref{rho_ent_coherent}). In this case the
separable and entangled photons induce different correlations
$R^{\rm (c)}$ between the Josephson currents.

In figure 16 we present (a) $\langle I_{\rm A}\rangle _{\rm sep} -
\langle I_{\rm A}\rangle _{\rm ent}$ and (b) $\langle I_{\rm
A}^{2}\rangle _{\rm sep} - \langle I_{\rm A}^{2} \rangle _{\rm ent}$
corresponding to irradiation with separable and entangled coherent
states of equations (\ref{rho_sep_coherent}) and
(\ref{rho_ent_coherent}).

In figure 17 we show the $\langle I_{\rm A} I_{\rm B} \rangle _{\rm
sep} - \langle I_{\rm A} I_{\rm B} \rangle _{\rm ent}$ that are
induced by (a) the separable and entangled number states of
equations (\ref{rho_sep_num}), (\ref{rho_ent_num}); and (b) the
separable and entangled coherent states of equations
(\ref{rho_sep_coherent}), (\ref{rho_ent_coherent}).

In figure 18 we plot the difference $R^{\rm (c2)}_{\rm sep} - R^{\rm
(c2)}_{\rm ent}$ corresponding to (a) the separable and entangled
number states of equations (\ref{rho_sep_num}), (\ref{rho_ent_num});
and (b) the separable and entangled coherent states of equations
(\ref{rho_sep_coherent}), (\ref{rho_ent_coherent}).

%===============================================================================================================%
%------------------------------------ *** SECTION 9 *** --------------------------------------------------------9
\section{Discussion}
We have studied electron interference in mesoscopic devices in the
presence of nonclassical electromagnetic fields. The phase factor is
in this case a quantum mechanical operator, whose expectation value
with respect to the density matrix of the electromagnetic field
determines the electron interference. We have presented various
examples, which show that the quantum noise of the photons destroys
slightly the electron interference fringes. Related is also the fact
that the photon statistics affects the interfering electrons. These
ideas have also been applied in the context of mesoscopic SQUID
rings.

In certain cases we get novel quantum phenomena with no classical
analogue. For example, in the case of a mesoscopic SQUID ring
irradiated with microwaves in a squeezed vacuum state we get Shapiro
steps only at even multiples of the fundamental frequency.

An important feature of nonclassical electromagnetic fields is
entanglement. We have considered two distant mesoscopic electron
interference devices that are irradiated with a two-mode
nonclassical electromagnetic field. Each field mode is coupled to
one of the mesoscopic devices. For entangled electromagnetic fields,
the electric currents and their higher moments become correlated.

All our results have been derived within the external field
approximation where the back reaction (additional flux created by
the electrons) is negligible. This is a valid approximation in
devices with small inductance.

Most of the experimental work on mesoscopic devices has studied
their interaction with classical electromagnetic fields, until
recently \cite{experiment}. Our results show that there is merit in
having a full quantum system where both the mesoscopic device and
the electromagnetic field are quantum mechanical. In this case we
can have purely quantum phenomena, without classical analogue, such
as the entanglement of distant mesoscopic devices.

%===============================================================================================================%
%---------------------------------------------------------------------------------------------------------------%
%-------------------------------------- REFERENCES -------------------------------------------------------------%

%--------------------------------------------------------------------------------------------------------------%
\end{document}